\documentclass[sigconf]{acmart}
\usepackage[utf8]{inputenc}
\usepackage{tabularx,acmart-taps}
\newcommand{\add}[1]{#1}
\usepackage{booktabs}
\usepackage[autostyle=false, style=english]{csquotes}
\usepackage{array}
\usepackage{pgfplots}
\usepackage{xcolor,colortbl}
\usepackage{float}
\usepackage{caption}
\usepackage{tikz}
\usepackage{soul}
\usepackage{cellspace}
\usepackage{subcaption}
\usepackage{pifont}
\usepackage{svg}
\usetikzlibrary{arrows.meta}
\usetikzlibrary{arrows.meta,positioning,calc}
\pgfplotsset{compat=1.18}
\newcolumntype{Y}{>{\raggedright\arraybackslash}X}

\newcolumntype{L}[1]{>{\raggedright\arraybackslash}p{#1}}
\usepackage{array}
 
\usepackage{makecell}

\usepackage{cellspace}
\usepackage{pifont}
\usepackage{placeins} 
\usepackage{caption}
\AtBeginDocument{%
  \providecommand\BibTeX{{%
    \normalfont B\kern-0.5em{\scshape i\kern-0.25em b}\kern-0.8em\TeX}}}

\copyrightyear{2026}
\acmYear{2026}
\acmConference[CHI '26]{Proceedings of the 2026 CHI Conference on Human Factors in Computing Systems}{April 13--17, 2026}{Barcelona, Spain}
\acmBooktitle{Proceedings of the 2026 CHI Conference on Human Factors in Computing Systems (CHI '26), April 13--17, 2026, Barcelona, Spain}
\acmDOI{10.1145/3772318.3790828}
\acmISBN{979-8-4007-2278-3/2026/04}

\begin{document}

\author{Aditya Kumar Purohit}
\authornote{Both authors contributed equally to this research.}
\email{aditya.purohit@cais-research.de}
\affiliation{
  \institution{Center for Advanced Internet Studies}
  \city{Bochum}
  \country{Germany}
}

\author{Aditya Upadhyaya}
\email{aditya.upadhyaya@uni-wuerzburg.de}
\authornotemark[1]
\affiliation{
  \institution{University of Würzburg}
  \city{Würzburg}
  \country{Germany}
}

\author{Nicolas Ruiz}
\email{nicolas.ruiz@cais-research.de}
\affiliation{
  \institution{Center for Advanced Internet Studies
  \& University of Bremen}
  \city{Bochum}
  \country{Germany}
}

\author{Alberto Monge Roffarello}
\email{alberto.monge@polito.it}
\affiliation{
  \institution{Politecnico di Torino}
  \city{Turin}
  \country{Italy}
}

\author{Hendrik Heuer}
\email{hendrik.heuer@cais-research.de}
\affiliation{
  \institution{Center for Advanced Internet Studies
\& University of Wuppertal}
  \city{Bochum}
  \country{Germany}
}

\title[Graphonymous Interaction]{When Handwriting Goes Social: Creativity, Anonymity, and Communication in Graphonymous Online Spaces}

\begin{abstract}

While most digital communication platforms rely on text, relatively little research has examined how users engage through handwriting and drawing in anonymous, collaborative environments. We introduce Graphonymous Interaction, a form of communication where users interact anonymously via handwriting and drawing. Our study analyzed over 600 canvas pages from the Graphonymous Online Space (GOS) CollaNote and conducted interviews with 20 users. Additionally, we examined 70 minutes of real-time GOS sessions using Conversation Analysis and Multimodal Discourse Analysis. Findings reveal that Graphonymous Interaction fosters artistic expression, intellectual engagement, sharing and supporting, and social connection. Notably, anonymity coexisted with moments of recognition through graphological identification. Distinct conversational strategies also emerged, which allow smoother exchanges and fewer conversational repairs compared to text-based communication. This study contributes to understanding Graphonymous Interaction and Online Spaces, offering insights into designing platforms that support creative and socially engaging forms of communication beyond text. 

\end{abstract}



\begin{teaserfigure}
\centering
  \includegraphics[scale=0.22]{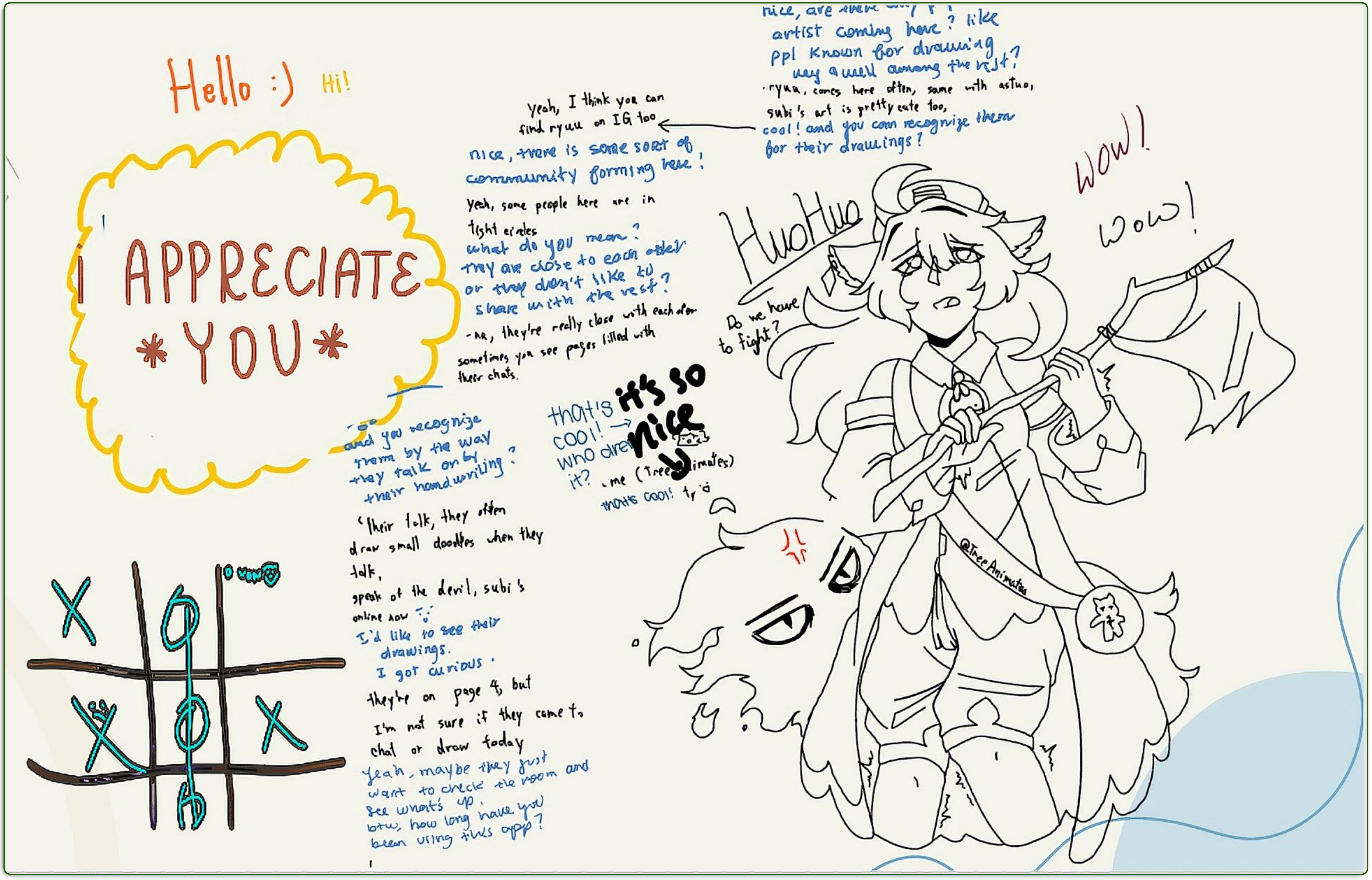}
  \caption{A creative conversation unfolds through digital handwriting and drawing, blending art with thoughtful interactions.}
  \Description{Anonymous user engagement  on a public doodle platform}
  \label{fig:teaser}
\end{teaserfigure}

\maketitle

\section{Introduction}

\add{Humans exhibit a tendency to draw from early childhood~\cite{Martinet2021,Fan2023}; however, these natural abilities have been largely absent from digital communication platforms.} In today's age, the nature of our communication with platforms like WhatsApp, Instagram, and Facebook is dominated by text-based communication via touch typing on smartphones and tablets. In research, much attention has been given to studying social interactions and conversational dynamics in text-based communication, such as instant messaging~\cite{crystal2009txtng,snapchattext,  greenfield2003online, upadhyay_aphasia}. However, research on alternative forms of digital communication using digital handwriting and drawing remains limited.

This paper explores an alternative and emerging form of communication and social interaction that we term \textit{\textbf{Graphonymous Interaction}} (henceforth, \textbf{GI}). GI refers to anonymous online interactions that take place through digital handwriting and drawing (see Figure~\ref{fig:teaser}), within a collaborative digital environment, which we call \textit{\textbf{Graphonymous Online Space}} (henceforth, \textbf{GOS}). This mode of interaction, which emphasizes creativity and non-verbal expression, has been relatively underexplored, despite its potential to enrich our social interactions. By shifting the focus from touch typing to a more expressive, visually creative modality, we aim to examine how creativity takes shape on GOS and how it may encourage deeper engagement in social media environments. Specifically, we define the following research questions:

\begin{itemize}
    \item RQ1: How does GOS foster social engagement and creativity?
    \item RQ2: What key conversational dynamics emerge in GI?
\end{itemize}

To explore these questions, we conducted an in-depth analysis of user behavior and interactions within a specific GOS called CollaNote~\cite{collanote2024}. Our methodology included a thematic analysis of more than 600 digital canvas pages from the GOS, as well as interviews conducted on 20 users on the same platform. We used a tablet and a digital stylus to facilitate digital handwriting and drawing, which allowed us to gather insights in an authentic environment. Furthermore, we conducted Conversation Analysis~\cite{tan2006conversational} (henceforth, CA) and Multimodal Discourse Analysis~\cite{MDA2021} (henceforth, MDA) to examine real-time Graphonymous Interactions on the GOS and understand how participants systematize multi-modal interactions through spatial and visual turn-taking, in contrast to linear text-based chats. 

Our findings indicate that the GOS fosters engagement and creativity through artistic expression, intellectual engagement, sharing and supporting, graphological identification, and social connection. Interestingly, despite the platform’s promotion of anonymity, users reliably identify each other based on their digital handwriting. CA and MDA of GIs reveal that users are not just participating in conversations, but are also actively creating a new register: a language variety used within a particular communicative context that blends text, visual, and paralinguistic cues for more creative and dynamic communication. 
Unlike text-based platforms, where linear turn-taking can cause overlapping messages and topic decay, GIs facilitate smoother conversational flow through a non-linear layout. Users employ visual markers, such as arrows to link utterances or separate texts, creatively structuring their real-time interactions. However, these workarounds highlight the limited threading tools in the GOS, which can make the process somewhat cumbersome. Our analysis of GIs also reveals minimal conversational gaps and overlaps, as users actively monitor each other’s writing in real-time, resulting in fewer conversational repairs. These insights suggest that GI offers a promising new modality for richer and more creative social interactions, despite the platform's design constraints.

This study contributes to the literature by identifying a novel form of online communication, thereby offering a distinctive alternative to conventional identity and text-centric platforms. This study makes three key contributions.

(1) To our knowledge, this is the first empirical study to introduce and define \textit{Graphonymous Interaction} (GI) as a novel communicative form embedded in anonymous, collaborative handwriting and drawing environments. We conceptualize GOS as a new sociotechnical space that supports such interactions, thereby highlighting its emphasis on visual expressiveness and a non-linear conversational flow.

(2) Through a thematic analysis of over 600 canvases from the GOS and accompanying user interviews, we present novel perspectives on how handwriting facilitates creativity, emotional expression, intellectual engagement, and user identification, even in the face of the platform’s anonymity. We also demonstrate how users collaboratively construct a new communicative register that seamlessly blends text and paralinguistic elements, resulting in rich and multifaceted social interactions.

(3) We explore the potential applications of graphonymous interaction (GI) and graphological output (GOS) across various domains, such as artistic democratisation, mental health support, community building, and alternative social media design. In doing so, we also identify the critical challenges and risks associated with this interactional mode, including pseudo-anonymity and perceived safety, authorship ambiguity, and interactional complexity stemming from user-devised adaptive strategies.  Finally, we reflect on the limitations of our study and outline promising directions for future research on graphonymous interactions.

\section{Background and Related Work}
\label{sec:related}

In this section, we examine the forms of use of handwriting and drawing in practice, focusing on digital mediums. While research on digital handwriting and drawing for communication is scarce, we situate our findings with respect to digital inking. We also review creativity and self-expression in digital communication, as well as conversational dynamics in digital spaces. This will provide a foundation for understanding the broader implications of our research.

\subsection{Digital Handwriting and Drawing in Practice}

In our research, we use the term “digital handwriting and drawing” to emphasize the personal, social, and creative aspects of writing and drawing on digital surfaces such as tablets with digital pens. Previous studies often refer to this practice as “digital inking” to describe the technical process of writing and drawing with a stylus~\cite{digitalinking}.  Previous research on digital inking has explored its applications in digital note-taking~\cite{styleblink,livesketch}, personal reflection~\cite{activeink}, learning~\cite{tablet_comp,khan-tablet}, and active reading~\cite{liquidtext}. Notably, digital note-taking research highlights how users organize their thoughts during activities such as brainstorming and journaling, frequently using lists, tables, calendars, and diagrams to structure their ideas more effectively on a digital canvas~\cite{styleblink}. In 2024, Helen et al.~\cite{graphicallyspeaking} used autoethnography to explore how scholars use digital and analog note-taking to enhance their understanding and recall of academic material. Researchers have explored the use of digital inking beyond productivity and learning aspects. For instance, digital inking has been shown to aid bipolar patients in exchanging deeper truths about their mental health~\cite{syndersmentalhealth}. 
This underscores the fact that the communicative role remains largely unexplored.

\add{To enhance learning with tablets}, researchers have developed applications like InkChat to facilitate learning, especially in mathematics, where formulas are difficult to express using a keyboard or mouse but can easily be conveyed through handwriting~\cite{Smith2013}. The primary goal of InkChat was to enable cross-device collaboration using a stylus on a shared canvas, although the research focused primarily on tool development without a deeper analysis of interaction dynamics. There is evidence that digital handwriting can improve recall~\cite{tablet_comp}, student performance, and engagement~\cite{khan-tablet,typerighting,stickel2008lessons} compared to traditional touch-typed text. The increasing adoption of tablets and styluses in classrooms is not surprising, as handwriting, often combined with drawing, naturally enhances expression, and improves teacher-student interactions~\cite{10.1145/1753846.1754214}. Similarly, the use of digital drawings has surged, particularly in design and creative fields. Recent developments include SketchPath, which uses hand-drawn toolpaths on a digital canvas for clay 3D printing~\cite{frost2024sketchpath}, and InkBrush, which enables free-form 3D ink stroke creation for digital painting~\cite{inkbrush}. 

\add{While these developments demonstrate the versatility of digital handwriting and drawing across domains, collaborative platforms employing these modalities remain relatively understudied from an interactional and social perspective. Only a few collaborative drawing tools, such as FlockMod and HelloPaint, exist commercially. However, CollaNote's public room design differs fundamentally in ways that make it ideal for studying Graphonymous Interactions (GI) as defined in this work.}~\add{FlockMod~\cite{FlockMod2024} is a browser-based tool that enables users to draw together online with optional account registration, operating through room-based structures where registered users maintain persistent usernames visible in moderation logs and user lists, a feature that contradicts core GOS principles of visual anonymity. Room creators establish control over access rules which introduces hierarchical governance that diverges from egalitarian community moderation. HelloPaint~\cite{HelloPaint2024}, an early-beta, browser-based collaborative painting platform focused on a global artist community, offers public galleries with persistent handles and portfolios, which grounds user identity in stable profiles rather than the anonymity fundamental to GI. Critically, both platforms employ persistent canvas storage. While this is not a definitional requirement for GOS, it, nevertheless, supports ongoing identity accumulation and reduces the low-stakes nature of anonymous exchange.}

\add{In contrast, the GOS examined in this study embodies fundamental GI design principles: contributions are visually anonymous, with no username attribution on the canvas itself, which ensures pseudo-anonymity, community-wide moderation, rather than control by creators or registered users, ensuring egalitarian governance, and participants must communicate through handwriting and drawing, rather than relying on text chat in sidebars or username-based attribution. CollaNote's specific implementation includes hourly canvas resets, a design choice that enforces ephemerality and reduces concerns about long-term identification, thereby allowing more candid self-expression and creative experimentation. Nonetheless, this reset mechanism is a unique aspect of CollaNote's GOS design and not a required feature of all GOS or GI environments. The fundamental defining characteristics of GOS, which include visual anonymity and a focus on handwriting-based communication, continue to be the key elements that set GOS platforms apart from traditional text-based or identity linked alternatives.}

\add{Despite extensive research on digital handwriting for learning and note-taking, understanding how they enable anonymous, real-time social communication remains largely unexplored. Our research introduces Graphonymous Interaction, where anonymity and visual expressiveness converge to foster creative exchange. We examine how GOS design principles facilitate communication without the identity accountability that often constrains expression on conventional social platforms.}

\subsection{Creativity and Self-expression in Online Social Networks}

Today, social media platforms are equipping users with tools for creative expression and production. While concerns exist regarding the potentially harmful and addictive effects of social media use~\cite{Baughan2022,Zhang2022}, researchers have also recognized its positive impact on creativity. Social media users can consume and create content while collaborating with others online~\cite{kaplan2010users}. The ease of sharing ideas and expressing oneself has effectively ``democratized'' creativity, making it accessible to a wider audience~\cite{creativity_sm,allen2012creativity}. Social media platforms also enable users to receive feedback on their ideas~\cite{creativity_sm}. For example, sharing creative content on Instagram exposes it to a large, unpredictable audience~\cite{reel_insta}, reflecting a willingness to take social risks, which is critical for fostering personal creativity~\cite{creativity_risks}.

Recent research by Acar et al.\cite{creativity_sm} highlights that active engagement on social media, such as posting and sharing content, is more strongly linked to creativity than passive activities such as browsing others' posts. Additionally, the relationship between creativity and social media use is more pronounced on platforms such as X (formerly Twitter) than on Instagram~\cite{creativity_sm}. However, research has also found that social media, particularly Instagram, can negatively affect creativity by fostering anxiety and social comparison~\cite{seabra2021art}. The literature discussed thus far emphasizes the role of active engagement in fostering creativity, highlights how social comparison can inhibit it, and underscores the importance of interaction dynamics that can either enhance or limit creative expression. This directly informs our exploration of digital handwriting and drawing in the GOS.

Creativity in social networks often flourishes through increased visibility and feedback. While anonymity is frequently associated with trolling and hate speech, it also offers users a unique space to express themselves freely without fear of judgment~\cite{strangersonphone}. This is especially valuable in settings where individuals may feel silenced or marginalized. Studies of anonymous platforms illustrate the complex dynamics of anonymity in real-world settings. These platforms allow users to share ideas and content without revealing their identities, often leading to candid, creative, and sometimes controversial expressions ~\cite{doi:10.1177/14614448221078603}. For example, Yik Yak, an anonymous location-based social app, gained popularity among college students by facilitating open discussions and building a sense of community. However, it struggled with issues such as harmful content and cyberbullying, which ultimately contributed to its decline~\cite{yikyak}. These examples highlight the dual-edged nature of anonymity by showing how the freedom to express oneself can have negative consequences. Striking a balance between anonymity and preventing misuse is crucial for fully realizing the benefits of online social networks~\cite{needtotalk}. This balance motivated us to study the GOS and GI, where we aim to explore how digital handwriting and drawing can foster creativity and collaboration in anonymous settings.

\subsection{Conversational Dynamics in Digital Spaces} \label{relwork}

Dynamics of face-to-face social interactions have long been studied using CA, which is a qualitative
research method that ``examine[s] how human beings use language to communicate'' ~\cite{liddicoat2021introduction} and interact socially, both verbally
and non-verbally. It allows the study of users' linguistic and interactional practices, such as how they organize their talks and deal with the problems they encounter during interactions~\cite{egbert2012introduction}. To better understand these conversational dynamics, CA examines several key components that shape the structure and flow of the conversation. These practices include the following:

 \begin{enumerate}
\item \textbf{Turn-taking:} refers to how participants organize the exchange of speaking turns during conversation, which is not a random phenomenon. Turn-taking functions effectively by allowing the production of ``long stretches of turns-at-a-talk'', which follow one another with minimal gaps and overlaps ~\cite{schelgoff2007sequence}. This concept is based on three principles. First, one speaker typically speaks at a time in a meeting. Second, turns are made up of turn constructional units (TCUs), which can be words, phrases, or clauses. A turn generally ends when a TCU is complete, creating a natural point where another speaker might take over, known as the transition relevance place(TRP). Third, at a TRP, turn allocation is used to decide how the next speaker is selected (other/self-selection). 

\item \textbf{Sequence Organisation:} refers to the way conversational turns are linked to form coherent and meaningful actions~\cite{schelgoff2007sequence}. The concept involves the idea of basic building blocks called adjacency pairs, which are closely connected to turn-taking. Adjacency pairs consist of two-part exchanges produced by different speakers, where the first part prompts a specific type of response in the second part, such as greeting-greeting, question-answer, invitation-acceptance/refusal, etc. ~\cite{levinson1983pragmatics}. The phenomenon that links the first and second part is called conditional relevance, which means the relevance of the second actions is dependent upon the occurrence of the first ~\cite{schelgoff2007sequence}. Conditional relevance guides turn-taking by signaling when a response is required. For instance, a request for information creates an expectation for a reply, which helps shape the structure of the conversation. 

\item \textbf{Repair:} refers to the process of identifying and correcting problems in speaking, hearing or understanding by participants. Repairs are crucial for preserving the flow and coherence of conversation. Sometimes, repair is initiated by the same speaker (self-repair) when they notice a problem in their speech and correct it. Alternatively, repair can be initiated by one speaker and completed by the other speaker (other-repair) through asking for clarification, requesting repetition, or directly correcting what was said ~\cite{schelgoff2007sequence}. Repairs are crucial in conversation, as they help manage the flow of conversation by allowing participants to correct themselves or others without disrupting the flow.

\end{enumerate}

With the rise of online communication, several studies in recent times have extended CA beyond face-to-face interactions to text-based digital spaces where users primarily type/tap, revealing both similarities and differences between physical and virtual interactions ~\cite{meredith2014repair, al2023pragmatic, greenfield2003online, tan2006conversational, koivisto2023conversation}. For example, ~\citet{greenfield2003online} studied conversation dynamics in a teen chatroom and found that teenagers employ strategies familiar from face-to-face interactions, such as addressing participants by name and repeating parts of statements to organize their conversations. They also identified unique strategies such as typing specific numerals to engage on a specific topic with a specific participant to create coherence. Similarly, ~\citet{meredith2014repair}’s  study on Facebook chat shows that features like repair and laughter are not random but are “precision-timed and designed to accomplish action”. In another study on the instant messaging chat system Internet Relay Chat (IRC),~\citet{al2023pragmatic} found that turn-taking and adjacency pairs are often complex and ambiguous because of factors such as time lags in message delivery, multiple people typing simultaneously, and frequent topic shifts. A more recent study on Tinder analyzed the use of wink emojis in flirting interactions and found that emoji ambiguity arises from factors such as idiosyncratic use, sequential position, and textual interplay~\cite{gibson2024flirting}. This often leads users to gloss over potential ambiguities to keep their conversations aligned.
Despite these limitations and ambiguities, ~\citet{herring1999interactional} argues that computer-mediated communication (henceforth, CMC) remains popular due to participants’ ability to adapt well to the medium and enjoy increased interactivity and language play that comes with loosened coherence. However, while CA has been applied to text-based CMC, it has yet to be implemented to study real-time GI, particularly in multi-modal environments like GOS, where users engage in both drawing and chatting, using a digital stylus on a tablet. To bridge this gap, we expand the scope of CA by applying it to this novel in-the-moment interaction modality to investigate how users co-create and negotiate meaning in GIs. This method enables us to explore how conventional CA concepts described above emerge in a digital environment where spatial and visual cues significantly shape conversation dynamics.

To complement CA’s focus on GIs, this study also employs MDA, which focuses on the semiotic meaning of spatial and visual elements in communication ~\cite{van2005multimodality}. Precisely, MDA is an interdisciplinary domain that addresses how diverse semiotic resources come together to generate meaning in “multimodal phenomena" ~\cite{o2011multimodal}. Snyder describes it as the “study of the intersection and interdependence of various modalities of communication within a given context" ~\cite{snyder2010applying}. Its relevance emerges from the need to account for the complexity of modern human discourse, especially in digital media where text, symbols, images, and videos are inextricably linked ~\cite{chen2020visualizing, o2011multimodal}. Since meaning is seldom conveyed through verbal language alone, MDA, along with CA, becomes crucial for understanding such “semantic expansion" ~\cite{o2008systemic, chen2020visualizing}. Thus, in multi-modal environments like GOS, where users combine drawings and spatial cues with text, MDA can provide useful insights into how these elements convey interactive meaning and identity. However, despite MDA's vast adoption as a theoretical and analytical framework in education ~\cite{lim2021investigating, alyousef2016multimodal, alyousef2020sf}, social media and advertising ~\cite{eisenlauer2014facebook, ng2018critical, sari2021multimodal}, social and political discourse ~\cite{lennon2021multimodal, machin2016multimodality}, its application to dynamic interactions remains limited. While CA remains central for capturing conversational structures on the GOS, MDA supports this study by offering a more holistic understanding of real-time GIs.

\section{Methodology}
\label{study}
To address our research questions, we employed a multi-method qualitative approach. This involved thematic analysis~\cite{braun2006using}, semi-structured user interviews~\cite{sketch_dialogue}, and CA/MDA~\cite{schelgoff2007sequence, van2005multimodality} to investigate creative behaviors and interactions within the GOS. To reiterate, GOS refers to digital environments where users collaborate and interact anonymously through digital handwriting and drawing. Our dataset comprised over (a) 600 digital canvas pages, representing anonymous interactions between users via digital handwriting and drawing, analyzed through thematic analysis, (b) 20 semi-structured interviews with the GOS users, and (c) CA and MDA of 10 live GOS sessions, each lasting 5-10 minutes. The interviews were conducted on tablets using a digital stylus, which assisted communication through digital writing and drawing with the participants. 

\subsection{Features of the Graphonymous Online Space}
We selected CollaNote~\cite{collanote2024} as the GOS to run our analysis. CollaNote provides a collaborative space where users can interact on a shared digital canvas. Users have the ability to draw or write directly on the canvas using a stylus or finger. The space offers various customization options, such as adjustable pen styles and image insertion. Moreover, if users prefer not to use a digital stylus or finger for interaction, they can also interact via digital keyboard. The space operates under specific interaction guidelines that restrict long strokes to maintain a smooth user experience. 

Users must create an account to interact with the canvas, selecting a username during registration with the ability to edit it later. Although contributions on the canvas are anonymous, notifications temporarily appear to inform users whenever someone has drawn or erased content on a particular canvas page. These notifications are designed to be non-intrusive to maintain the flow of interaction. The GOS operates under well-defined rules, which prohibit erasing others' content, using profanity, and engaging in inappropriate behavior. These rules are prominently displayed within the interface to promote a respectful and collaborative environment, with active moderation by designated individuals to ensure compliance. The GOS resets hourly, refreshing the canvas for new interactions. Additionally, it offers an incentive system where users whose drawings are favored by the moderator are labeled as ``artists''. This designation grants them exclusive access to draw or write on the main page. \add{To use this specific GOS, users need to have the basic skill of operating a stylus or input device for writing or drawing. Based on our subjective perception, the GOS we observed is accessible to everyone from beginners to advanced artists.}

\subsection{Ethical Considerations}

This study underwent review and approval by the responsible authorities, who granted IRB-equivalent approval. Before participating in the semi-structured interviews on the GOS, all interview participants provided informed consent. Users’ identities on the platform were anonymised, and all visual data used in the study was publicly accessible. Permission was obtained from the team of Collanote over email before collecting any data. To maintain user anonymity and adhere to ethical standards, we have removed or altered any potentially identifying visual content in the results.

\begin{figure*}[tbp]
\centering
  \includegraphics[width=0.95\textwidth]{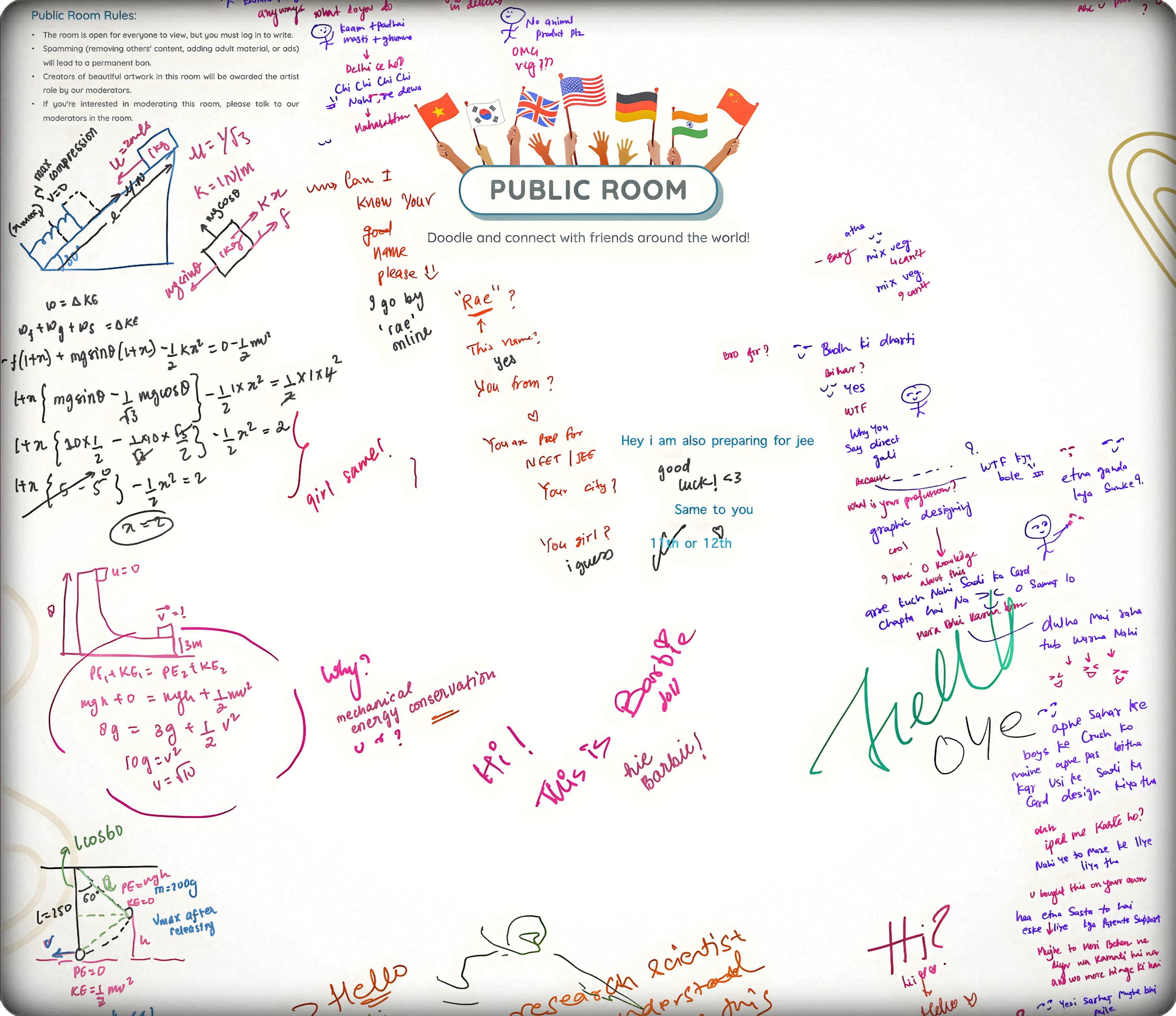}
  \caption{An example of a conversation extracted from CollaNote and converted into PDF for our analysis.}
  \Description{Anonymous user engagement on a public doodle platform}
  \label{fig:GOS_snapshot}
\end{figure*}

\subsection{Procedure and Data Analysis} 

\subsubsection{Thematic Analysis on Digital Canvases}
We conducted a thematic analysis ~\cite{braun2006using} to identify emerging patterns in user behavior and interaction. The first, second, and third authors participated in the coding process and examined over 600 canvas pages from the selected GOS. We converted this lively and novel space into PDFs to archive and analyze it (see Figure~\ref{fig:GOS_snapshot}).

At any given time, the GOS has 40-60 users interacting in real-time, writing anywhere on the canvas. This lack of linear order in conversations, with interactions occurring simultaneously, posed methodological challenges. We addressed these by utilizing Miro boards for analysis, which allowed for better organization and visualization of the data. In the first phase of analysis, each researcher independently reviewed their assigned snapshot of canvases, took notes and generated initial codes using axial coding principles~\cite{corbin2014basics}. This process was conducted on the Miro board, and was informed by first impressions of user interactions on the digital canvases and by repeated reviews of participants’ responses. The focus was on familiarizing ourselves with the data and identifying potential codes for user behaviors and creative actions. After completing their independent coding work on assigned canvas snapshots, the first four authors met to conduct an initial comparison of their codes. These preliminary 
codes were then reviewed collaboratively with the fifth author during weekly meetings, where any discrepancies were deliberated until consensus was reached. In the second phase, the researchers applied the agreed-upon codes to each PDF sheet, which had been converted from the canvases, focusing on user behaviors and interactions across the GOS. 

\subsubsection{Interviews on the GOS}
\label{interviewsonGOS}
To gain deeper insights into users' experiences and behaviors, we conducted semi-structured written interviews with 20 participants who actively engaged on the platform. \add{Of the 20 interview participants, 15 (75\%) provided demographic information. All participants were asked about their demographics; however, many users value anonymity in this space, which may have limited disclosure. Among those who responded, 11 identified as female and four as male, with ages ranging from 18 to 29 years (M = 21.5, SD = 3.2). This suggests that the platform primarily attracts young adults, particularly young female users seeking creative expression in anonymous spaces.} The interviews took place within the same environment, using tablets and digital styluses. This provided participants with an authentic experience, allowing them to respond to prompts and questions in the same medium they used for daily interaction.

We started with questions investigating \textbf{\textit{motivations}} for use (\textit{How frequently do you visit this public room to read others’ conversations? How regularly do you use this room to participate actively? Why do you come to this public room?}). Then, we explored how users \textbf{\textit{managed their conversations}} on the platform (\textit{How do you start talking to someone? How do you show that you’re open to chat? How do you signal when you want to end a conversation?}), as well as how they \textbf{\textit{trust their interactions}} (\textit{Can you describe a scenario where you decided to trust another user, and what made you trust them? What was the most trustworthy interaction you had, and what made it trustworthy? What was the least trustworthy interaction you experienced, and what made it untrustworthy? How do you show that you trust someone? Are there any specific words or actions?}). Finally, we closed the interviews with a thought-provoking question by asking participants to \textbf{\textit{compare their experience}} on the current platform with those on other social media platforms (\textit{What makes this public room better compared to other social media platforms?}). 

\subsubsection{Conversation Analysis and Mutlimodal Discourse Analysis} 
In addition to thematic analysis and interviews, we conducted CA and MDA exclusively on real-time GIs on the GOS to capture spontaneous, handwritten exchanges among anonymous users. It was conducted on 10 GOS sessions (5–10 minutes each, totaling 70 minutes) held on different days. Drawing on CA literature, we focused specifically on three key areas within CA: turn-taking, sequence organization and repair phenomenon (see ~\ref{relwork}). Additionally, MDA was employed to understand users' communication through visual and spatial collaboration. More specifically, the study adopts the SF-MDA (Systemic Functional Multimodal Discourse Analysis) framework, which examines how meaning is constructed through the combined use of language and visual imagery. This includes analyzing linguistic and visual modes of semiosis (such as handwriting and color), and examining how these elements interact to create meaning, a process known as intersemiosis ~\cite{o2008systemic, o201413, halliday1978language}. We conducted MDA in three stages: (i) systematic coding of visual and spatial elements, (ii) identification of how these elements are mobilized in interactions on the GOS, and (iii) interpretation of multi-modal features with respect to communicative aim (e.g., collaboration or creativity). 
Furthermore, CA and MDA were carried out on chats that occurred exclusively in English, as the GOS allows participants from various countries to contribute and interact in multiple languages. Notably, a unique feature of the platform allowed us to view each stylus stroke made by users in real-time, thereby providing us with the precise view of their writing process and interaction dynamics.

\section{Results}

In the following section, we present the findings from the thematic analysis and interviews conducted to address RQ1 and CA/MDA used to explore RQ2. 

\subsection{Thematic Analaysis and Interviews (RQ1)}
\label{sec:thematic}

To explore social engagement and creativity within the GOS, thematic coding was applied to over 600 digital canvas pages extracted from the GOS, as well as to the 20 semi-structured user interviews conducted on the same platform. Six key themes emerged: artistic expression, feedback, intellectual engagement, sharing and support, graphological identification, and social connection. Each of the themes is discussed below in detail.

The analysis indicated that users of the GOS frequently engage in the creation and sharing of diverse artistic forms. During the coding process, the following common artistic styles were identified: cartoon, anime/manga, kawaii, and stick drawing, as observed in Table~\ref{tab:drawing_styles}. Notably, conversation style involving drawings accounted for 33.23\% of all interactions (see Table~\ref{tab:usage_conversational_style}). Users often used art to express emotions like empathy, activism, and frustration, which were frequently reflected in their drawings. Similarly, participants in the user interviews described the GOS as a shared creative space. For instance, P01 remarked, \textit{``People can draw like this together in one place! I don't think I have seen [anything like this]''}. Another user, P13, stated, \textit{``We draw and we talk''}.

\aptLtoX[graphic=no,type=html]{\begin{table}[tb]
    \centering
    \small
    \begin{tabular}{lrr}
      \toprule
      \textbf{Drawing Style} & \textbf{Count} & \textbf{Percentage} \\
      \midrule
      Cartoon & 293 & 40.47\% \\
      Anime/Manga & 159 & 21.96\% \\
      Kawaii & 100 & 13.8\%1 \\
      Stick Drawing & 67 & 9.25\% \\
      Sketch & 49 & 6.77\% \\
      Pictures & 28 & 3.87\% \\
      Silhouette & 12 & 1.66\% \\
      Pastel & 6 & 0.83\% \\
      Art Activism & 5 & 0.69\% \\
      Realistic & 4 & 0.55\% \\
      Watercolor & 1 & 0.14\% \\
      \midrule
      \textit{Total} & 724 & 100.00\% \\
      \bottomrule
    \multicolumn{3}{c}{\textbf{(a) Drawing styles.}}\\
    \end{tabular}\\
    \begin{tabular}{lrr}
      \toprule
      \textbf{Conversational Style} & \textbf{Count} & \textbf{Percentage} \\
      \midrule
      Without Drawings & 269 & 66.75\% \\
      With Drawings & 134 & 33.25\% \\
      \midrule
      \textit{Total} & 403 & 100.00\% \\
      \bottomrule
    \multicolumn{3}{c}{\textbf{(b) Conversational styles.}}\\
    \end{tabular}
  \caption{Distribution of drawing and conversational styles observed in GOS interactions. Subtable (a) reports drawing-style frequencies and subtable (b) reports whether interactions included drawings.}
  \label{tab:styles_combined}
    \label{tab:drawing_styles}
    \label{tab:usage_conversational_style}
\end{table}}{\begin{table}[tb]
  \centering
  \begin{subtable}[t]{\columnwidth}
    \centering
    \small
    \begin{tabular}{lrr}
      \toprule
      \textbf{Drawing Style} & \textbf{Count} & \textbf{Percentage} \\
      \midrule
      Cartoon & 293 & 40.47\% \\
      Anime/Manga & 159 & 21.96\% \\
      Kawaii & 100 & 13.8\%1 \\
      Stick Drawing & 67 & 9.25\% \\
      Sketch & 49 & 6.77\% \\
      Pictures & 28 & 3.87\% \\
      Silhouette & 12 & 1.66\% \\
      Pastel & 6 & 0.83\% \\
      Art Activism & 5 & 0.69\% \\
      Realistic & 4 & 0.55\% \\
      Watercolor & 1 & 0.14\% \\
      \midrule
      \textit{Total} & 724 & 100.00\% \\
      \bottomrule
    \end{tabular}
    \caption{Drawing styles.}
    \label{tab:drawing_styles}
  \end{subtable}

  \begin{subtable}[t]{\columnwidth}
    \centering
    \small
    \begin{tabular}{lrr}
      \toprule
      \textbf{Conversational Style} & \textbf{Count} & \textbf{Percentage} \\
      \midrule
      Without Drawings & 269 & 66.75\% \\
      With Drawings & 134 & 33.25\% \\
      \midrule
      \textit{Total} & 403 & 100.00\% \\
      \bottomrule
    \end{tabular}
    \caption{Conversational styles.}
    \label{tab:usage_conversational_style}
  \end{subtable}

  \caption{Distribution of drawing and conversational styles observed in GOS interactions. Subtable (a) reports drawing-style frequencies and subtable (b) reports whether interactions included drawings.}
  \label{tab:styles_combined}
\end{table}}

\begin{figure}[tb]
\centering
  \includegraphics[width=\columnwidth]{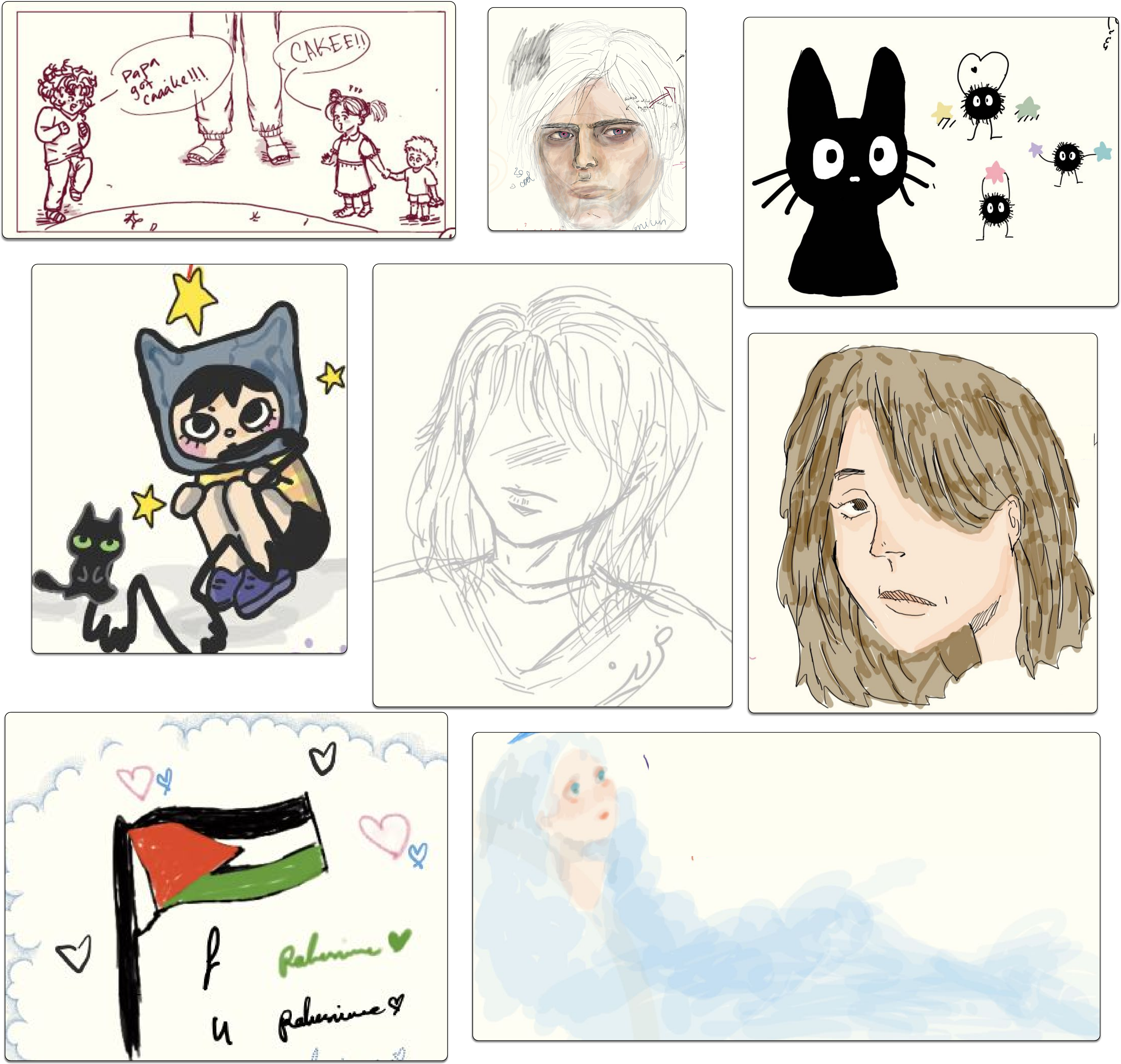}
  \caption{A collage of user-generated artwork from the GOS platform, showcasing a variety of artistic styles and expressions}
  \Description{A collage of user-generated artwork from the GOS platform, showcasing a variety of artistic styles and expressions.}
  \label{fig:GOS_artforms}
\end{figure}

The collage in Figure~\ref{fig:GOS_artforms} presents a diverse collection of artistic styles created by users on the GOS. In the top left, a comic-style sketch portrays a parent and children, while the top center features a realistic digital portrait with detailed facial features. The top right shows a minimalist cartoon illustration with a black cat and small creatures. The middle left showcases a kawaii-style drawing of a character wearing a cat hat alongside a small black cat. In the middle center, there is a lightly sketched portrait, giving an abstract and unfinished appearance, followed by a semi-realistic digital portrait on the middle right, utilizing soft shading and natural tones. The bottom left contains a simple illustration of the Palestinian flag, and the collage concludes with a watercolor-style painting of a mermaid in soft blue tones on the bottom right.

\begin{figure*}[tb]
\centering
  \includegraphics[width=\textwidth]{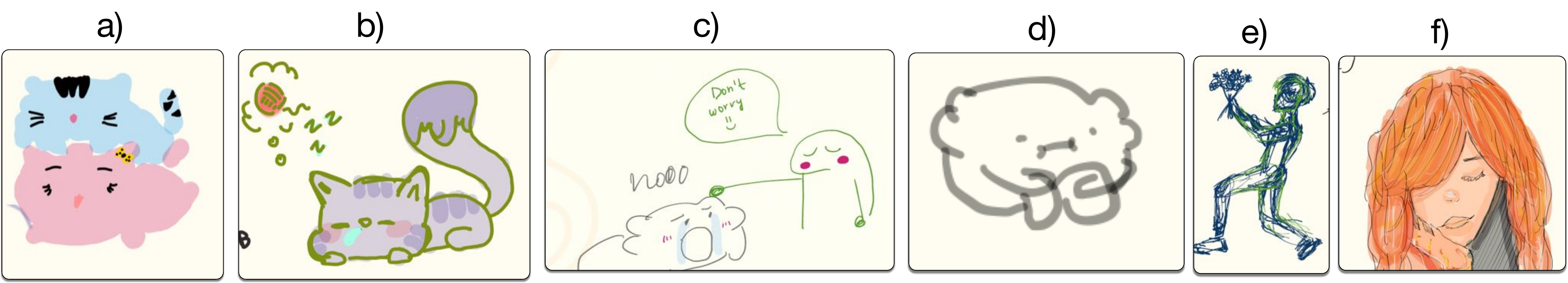}
  \caption{A collage of user-generated artwork on the GOS, showcasing a range of emotions through diverse artistic styles, including playfulness, calmness, empathy, love, and introspection }
  \Description{ A collage of user-generated artwork on the GOS platform, showcasing a range of emotions through diverse artistic styles, including playfulness, calmness, empathy, love, and introspection.}
  \label{fig:GOS_emotions}
\end{figure*}

Figure~\ref{fig:GOS_emotions} illustrates how users utilize the GOS to express their emotions through various artistic styles. The two colorful, cat-like creatures in (a) evoke a playful and cheerful mood, with soft shapes and pastel colors suggesting happiness and affection. The artwork in (b) shows a cat in relaxed postures, conveying calmness and tranquility. In (c), a blob-like creature comforts another with the words \textit{``Don’t worry =)''}, reflecting empathy and care. Figure~\ref{fig:GOS_emotions} (d) features a simple, cloud-like character with a neutral expression, suggesting introspection. The artwork in (e) presents an abstract figure in motion, with dynamic lines and poses that convey energy, and love. Lastly, (f) displays a detailed portrait of a person with their eyes closed, evoking serenity, and contemplation.

\subsubsection{Feedback}
\begin{figure}[tb]
\centering
  \includegraphics[width=0.5\textwidth,keepaspectratio]{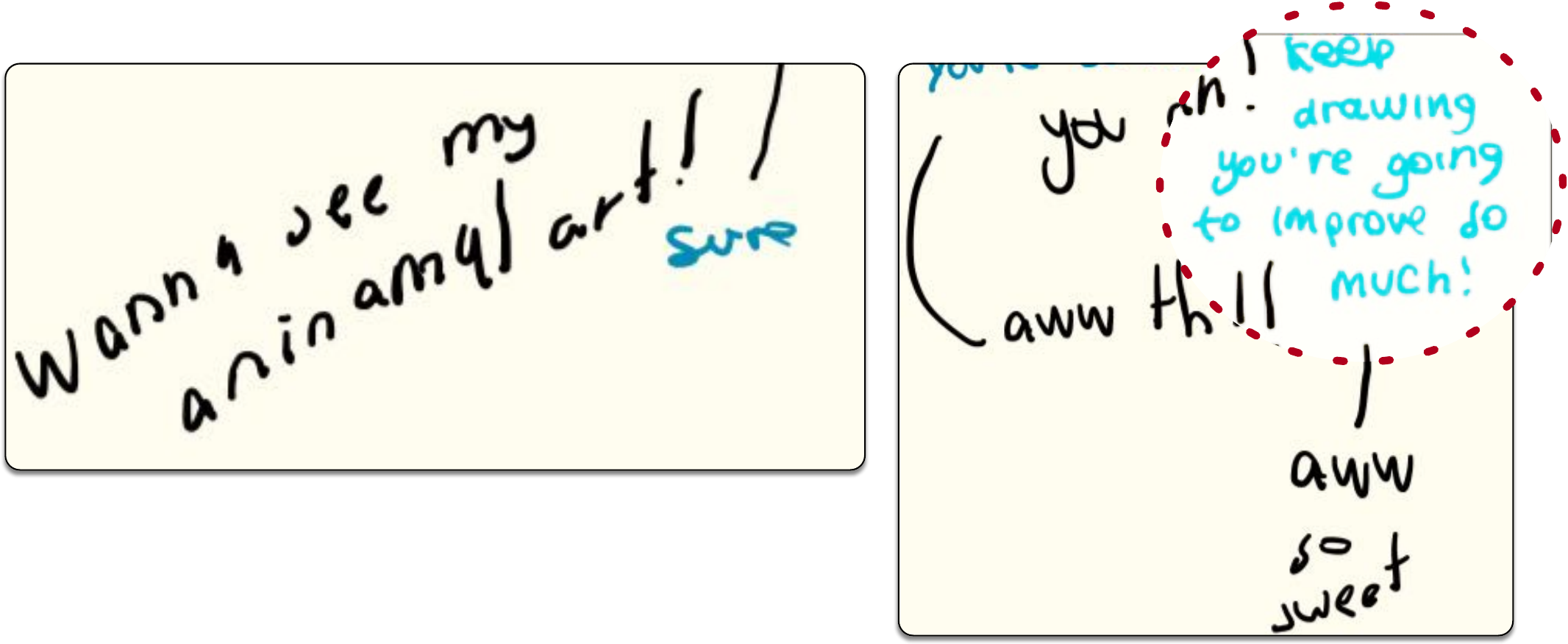}
  \caption{Example of user interaction on the GOS, where a user in black handwriting asks for feedback on their art, and another user in blue handwriting responds with appreciation and encouragement. }
  \Description{Example of user interaction on GOS, where a user in black handwriting asks for feedback on their art, and another user in blue handwriting responds with appreciation and encouragement}
  \label{fig:GOS_appreciate}
\end{figure}

\begin{figure*}[tb]
\centering
  \includegraphics[width=\textwidth,keepaspectratio]{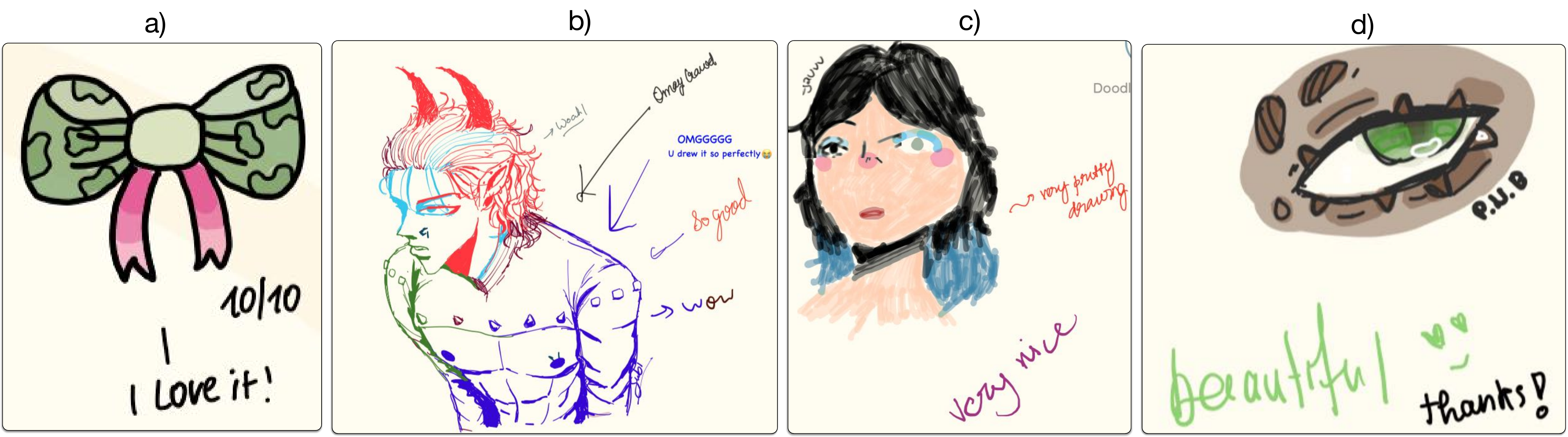}
  \caption{The collage highlights the collaborative environment fostered by spontaneous user feedback, and appreciation on art work. }
  \Description{The collage highlights the collaborative environment fostered by spontaneous user feedback and appreciation on art work.}
  \label{fig:GOS_feedback}
\end{figure*}

Another prominent theme was the role of feedback, appreciation, and advice in enhancing artistic expression. Artists frequently use the GOS to seek input on their draft work. In Figure~\ref{fig:GOS_appreciate}, we see an example where a user (in black handwriting) asks another user (in blue handwriting) if they would like to view their art. The user in blue not only appreciates the work but also encourages the artist, saying, \textit{``Keep drawing, you're going to improve so much !''}. In addition to direct requests, users often spontaneously comment on and provide feedback to others. For example, Figure~\ref{fig:GOS_feedback} (a) shows a user rating a piece 10/10 and expressing their love for the artwork. Other examples include an appreciation of a horned fantasy character illustration in Figure~\ref{fig:GOS_feedback} (b), a portrait of a girl in Figure~\ref{fig:GOS_feedback} (c), and an eye drawing in Figure~\ref{fig:GOS_feedback} (d).

\subsubsection{Intellectual Engagement}
\label{intellectengagement}

\begin{figure}[tb]
\centering
  \includegraphics[width=\columnwidth]{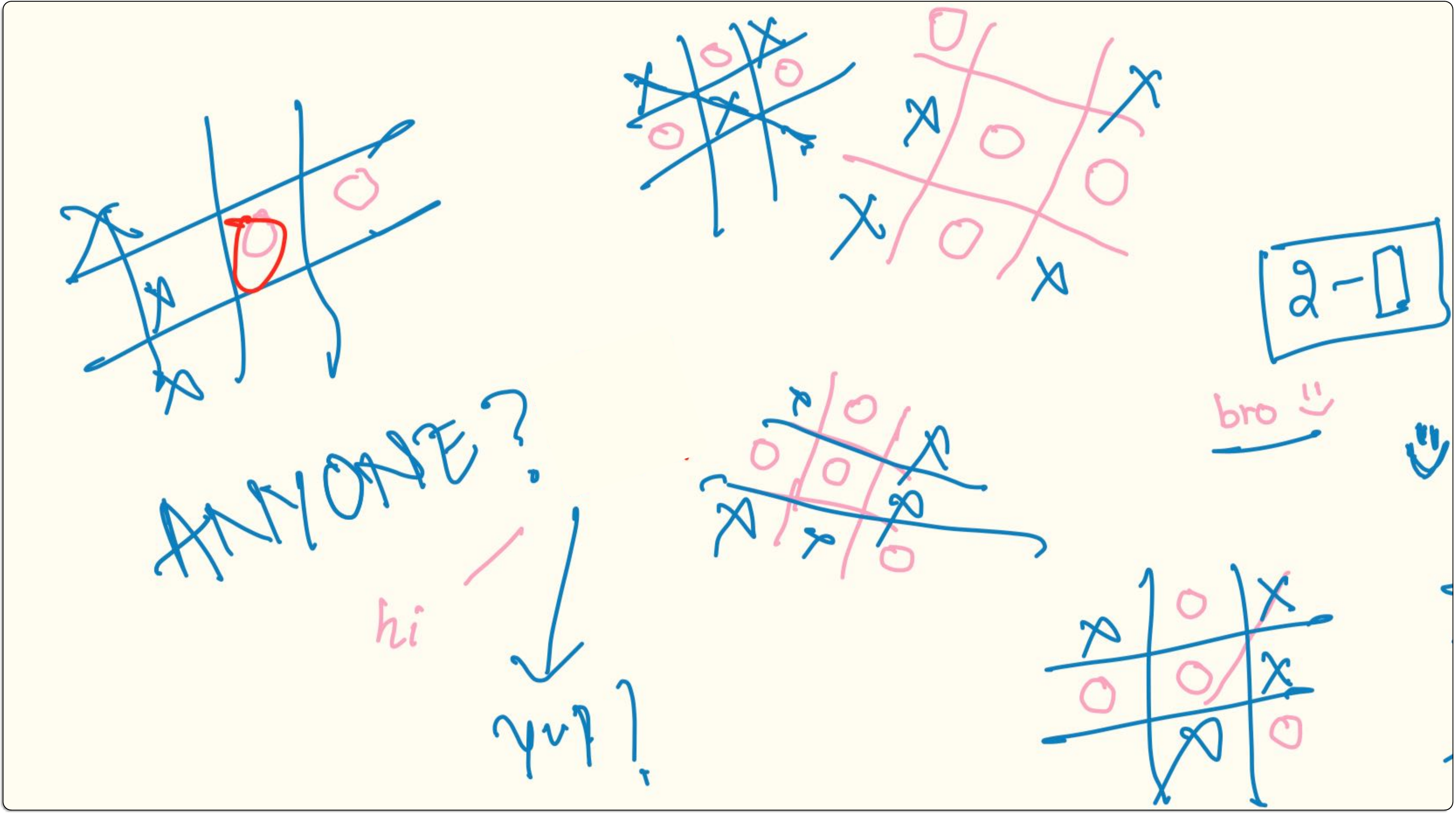}
  \caption{Collaborative Tic-Tac-Toe games on the GOS, with users engaging in real-time matches and inviting others to join.}
  \Description{Collaborative Tic-Tac-Toe games on GOS, with users engaging in real-time matches and inviting others to join.}
  \label{fig:GOS_intellectualengagement}
\end{figure}

The data revealed that users intellectually engage in different tasks on the GOS. Recurring codes that reflect this theme included abstract strategy games and math exercises. Users find creative ways to engage with each other. For instance, users draw games like Tic-Tac-Toe (two-player game played on a 3x3 grid) and invite others to join and play along. Among the games Tic-Tac-Toe is most played game (Table~\ref{tab:game_distribution}) and Math being the most discussed subject (Table~\ref{tab:subject_distribution}). In Figure~\ref{fig:GOS_intellectualengagement}, for example, a user in blue ink draws 3x3 grid and invites others by writing \textit{``ANYONE?''}. The user in pink handwriting responds with  \textit{``hi''}, then joins the game  and they keep track of their score as ``2-0''.

\aptLtoX[graphic=no,type=html]{
\begin{table}[tb]
  \centering
    \centering
    \small
    \begin{tabular}{lrr}
      \toprule
      \textbf{Game} & \textbf{Count} & \textbf{Percentage} \\
      \midrule
      TicTacToe & 45 & 95.74\% \\
      Hangman & 1 & 2.13\% \\
      Wordle & 1 & 2.13\% \\
      \midrule
      \textit{Total} & 47 & 100.00\% \\
      \bottomrule
 \multicolumn{3}{c}{\textbf{(a) Game usage distribution among GOS users.}}\\
   \end{tabular}
    \begin{tabular}{lrr}
      \toprule
      \textbf{Subject} & \textbf{Count} & \textbf{Percentage} \\
      \midrule
      Math & 11 & 42.31\% \\
      Physics & 5 & 19.23\% \\
      Biology/Medicine & 5 & 19.23\% \\
      Chemistry & 4 & 15.38\% \\
      Ethics & 1 & 3.85\% \\
      \midrule
      \textit{Total} & 26 & 100.00\% \\
      \bottomrule
\multicolumn{3}{c}{\textbf{(b) Distribution of subjects discussed on the GOS.}}\\
    \end{tabular}\end{table}}{\begin{table}[tb]
  \centering
  \begin{subtable}[t]{\columnwidth}
    \centering
    \small
    \begin{tabular}{lrr}
      \toprule
      \textbf{Game} & \textbf{Count} & \textbf{Percentage} \\
      \midrule
      TicTacToe & 45 & 95.74\% \\
      Hangman & 1 & 2.13\% \\
      Wordle & 1 & 2.13\% \\
      \midrule
      \textit{Total} & 47 & 100.00\% \\
      \bottomrule
    \end{tabular}
    \caption{Game usage distribution among GOS users.}
    \label{tab:game_distribution}
  \end{subtable}
  \begin{subtable}[tb]{\columnwidth}
    \centering
    \small
    \begin{tabular}{lrr}
      \toprule
      \textbf{Subject} & \textbf{Count} & \textbf{Percentage} \\
      \midrule
      Math & 11 & 42.31\% \\
      Physics & 5 & 19.23\% \\
      Biology/Medicine & 5 & 19.23\% \\
      Chemistry & 4 & 15.38\% \\
      Ethics & 1 & 3.85\% \\
      \midrule
      \textit{Total} & 26 & 100.00\% \\
      \bottomrule
    \end{tabular}
    \caption{Distribution of subjects discussed on the GOS.}
    \label{tab:subject_distribution}
  \end{subtable}
  \caption{Distributions of games and subjects observed in GOS interactions. Subtable (a) reports game usage and subtable (b) reports subjects discussed.}
  \label{tab:gos_distributions}
\end{table}}

\begin{figure*}[tb]
\centering
  \includegraphics[width=\textwidth,keepaspectratio]{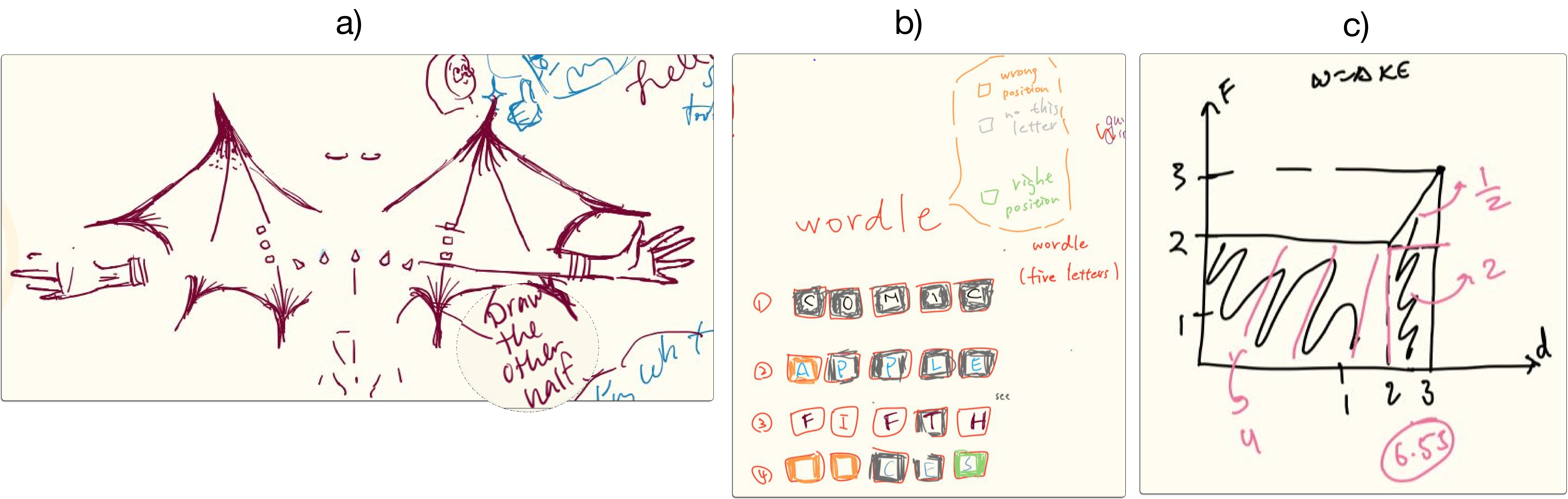}
  \caption{Examples of user collaboration on the GOS. (a) A half-finished drawing with the prompt ~\textit{“draw the other half”} (zoomed in for clarity) inviting creative contribution. (b) A collaborative Wordle game where users worked together to solve word puzzles. (c) A joint effort to solve a math problem, with users contributing to a graph and annotations. }
  \Description{Examples of user collaboration on the GOS. (a) A half-finished drawing with the prompt “draw the other half”, inviting creative contribution. (b) A collaborative Wordle game where users worked together to solve word puzzles. (c) A joint effort to solve a math problem, with users contributing to a graph and annotations.}
  \label{fig:GOS_art_math}
\end{figure*}

Similarly, users create partially finished drawings and invite others to complete them with prompts such as, \textit{``draw the other half''} (Figure~\ref{fig:GOS_art_math} (a)). This type of engagement is common on the GOS, and demonstrates how the platform encourages shared creative experiences. Collaborations extend beyond art as well, with users collaborating on activities like playing games such as Wordle (Figure~\ref{fig:GOS_art_math} (b)) and solving math problems (Figure~\ref{fig:GOS_art_math} (c)).

\subsubsection{Sharing and Support}
\label{sharingandsupportongos}

\begin{figure}[tb]
\centering
  \includegraphics[width=0.8\columnwidth]{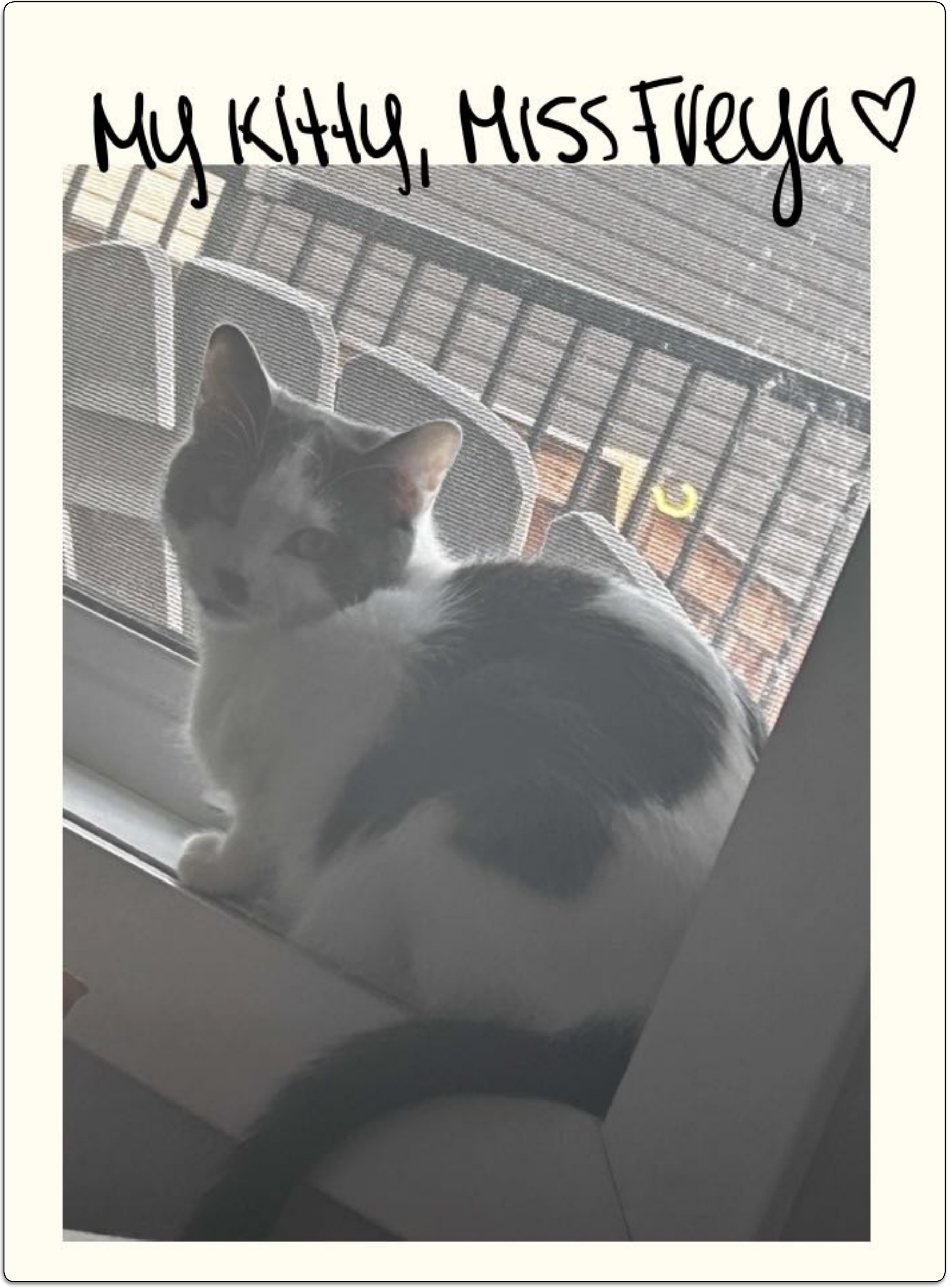}
  \caption{A user sharing a photo of their cat, Miss Freya, on the GOS. This highlights the platform’s use for sharing personal moments, such as pet photos, fostering a sense of community and connection. }
  \Description{A user sharing a photo of their cat, Miss Freya, on the GOS. This highlights the platform’s use for sharing personal moments, such as pet photos, fostering a sense of community and connection.}
  \label{fig:GOS_support_pet}
\end{figure}

\begin{figure}[tb]
\centering
  \includegraphics[width=\columnwidth]{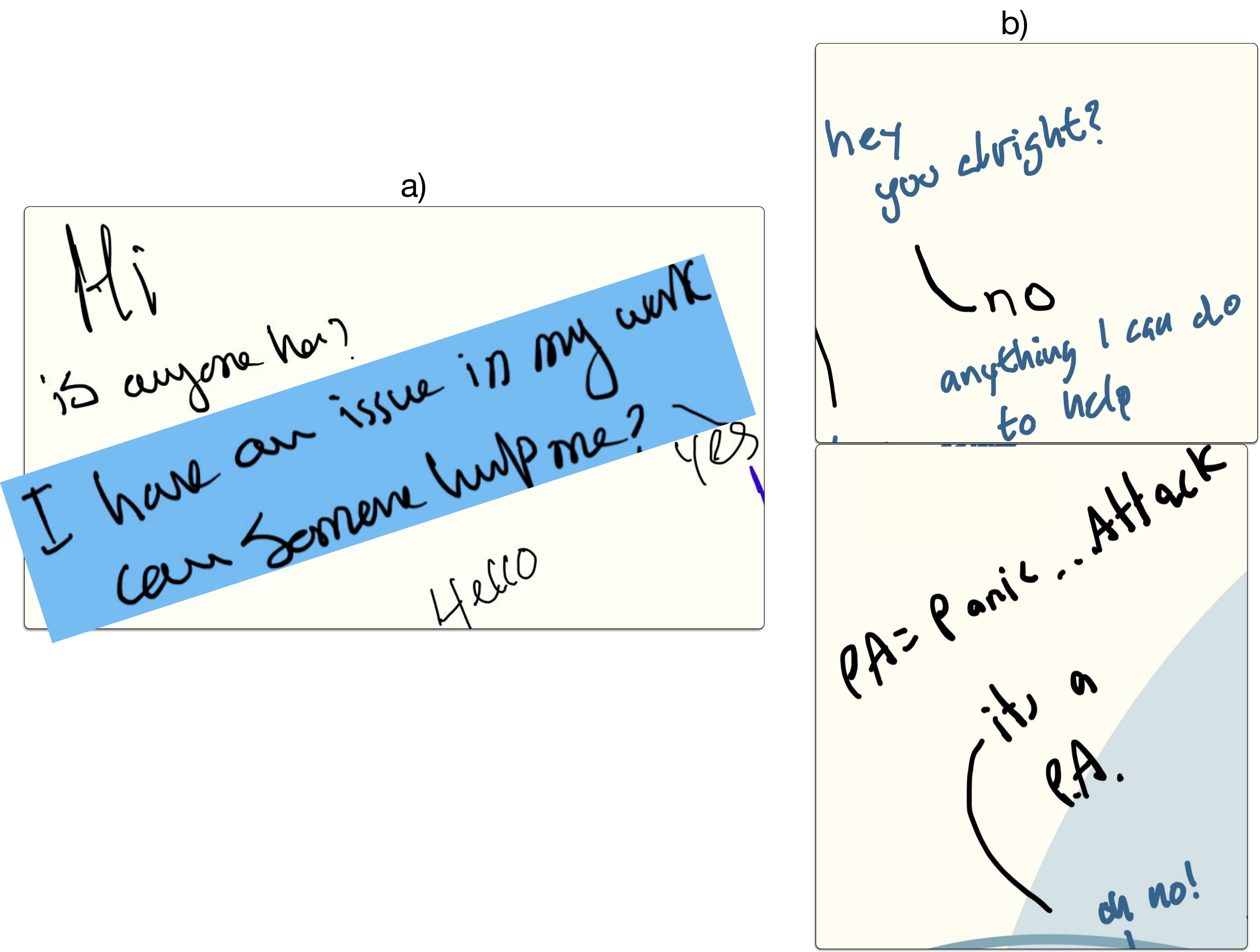}
  \caption{Examples of users seeking and offering support on the GOS. (a) shows a user asking for help with a work issue, while (b) depicts another user checking in on someone and offering assistance during a panic attack. This highlights the supportive nature of interactions on the platform, where users provide emotional and practical help. }
  \Description{Examples of users seeking and offering support on the GOS. (a) shows a user asking for help with a work issue, while (b) depicts another user checking in on someone and offering assistance during a panic attack. This highlights the supportive nature of interactions on the platform, where users provide emotional and practical help.}
  \label{fig:GOS_support_pain}
\end{figure}

The analysis revealed a recurring theme of sharing and supporting, where users frequently share personal experiences. Codes such as sharing pain, discussing social media, expressing fears, sharing pet photos (Figure~\ref{fig:GOS_support_pet}), and revealing emotions appear consistently throughout the data. For instance, one user wrote, \textit{``Can someone talk to me? I’m feeling depressed''}, while another expressed vulnerability by stating,  \textit{``I feel very embarrassed about my handwriting''}. In Figure~\ref{fig:GOS_support_pain} (a), a user seeks help with a work-related issue. Similarly, in Figure~\ref{fig:GOS_support_pain} (b), another user checks in on someone and offers support during a panic attack.

\begin{figure}[tb]
\centering
  \includegraphics[width=\columnwidth]{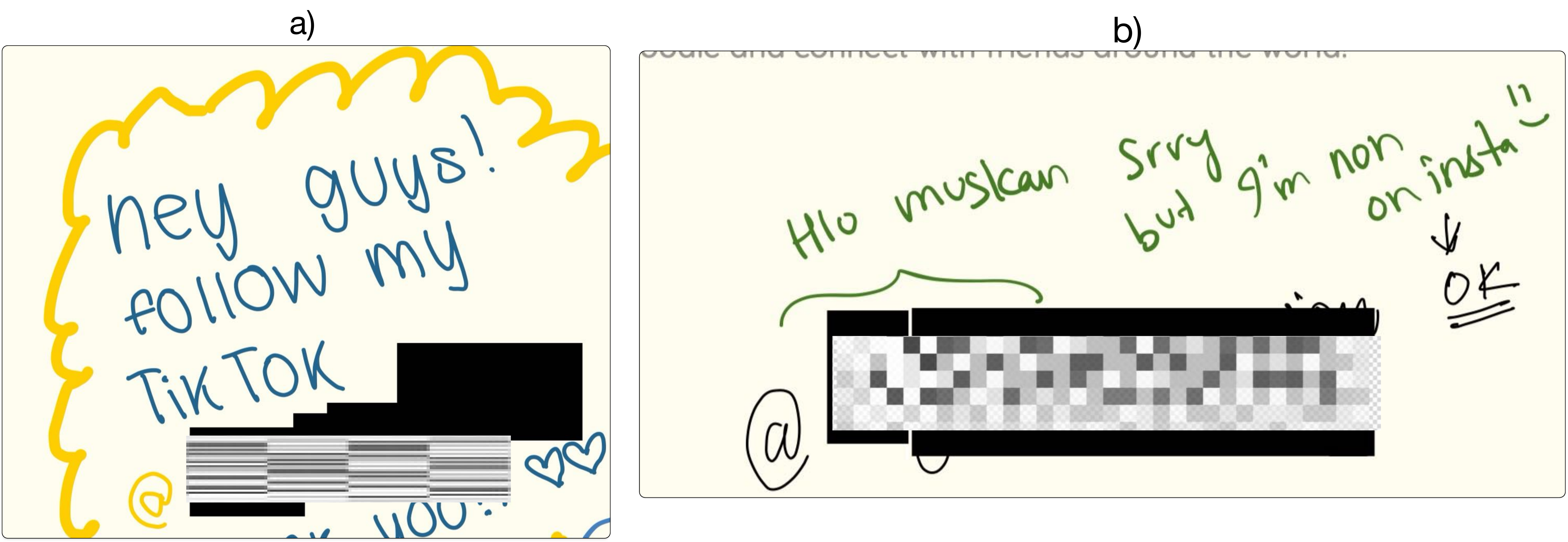}
  \caption{Redacted content for privacy reasons. Users sharing their social media handles on the GOS. (a) shows a user promoting their TikTok account, while  (b) displays users exchanging Instagram handles. }
  \Description{Users sharing their social media handles on the GOS. (a) shows a user promoting their TikTok account, while  (b) displays users exchanging Instagram handles} 
  \label{fig:GOS_support_socialmedia}
\end{figure}

We also observed attempts to form ongoing connections in other social media spaces. For example, one user encourages others to follow their TikTok account (Figure~\ref{fig:GOS_support_socialmedia} (a)), while other users exchange Instagram handles on the GOS to further expand their social networks (Figure~\ref{fig:GOS_support_socialmedia} (b)). 

In semi-structured interviews P19 wrote \textit{``I have dumped my emotions here   before and its better to do it respectfully''}. P18 wrote the following when asked if they found any support on the GOS \textit{``Sometimes I share my emotions here. e.g. when I had a bad day I can speak with some people here. And also it is comfortable here I don't feel like I have to show myself and my complete life''}. P07 reflected on the emotional support found on the platform, stating, \textit{``I talk to girls so they understand how I feel when I share about my life''}. This highlights the platform's role in fostering emotional support. A particularly revealing quote from P13, \textit{``I get panic attacks, bad mental health, I depend on [friends on this space] a lot''}, emphasizes that the platform allows users to self-report their emotions and the support they feel they receive.

Some participants (P13, P16, and P11) described the GOS as a community or family where they feel safe. For instance, P13 noted, \textit{``We have regulars and new people who come a few times and then disappear, but the regulars are like family, without judgment''}. Similarly, P16 shared, \textit{``I don’t know, I just feel like it’s safer here, it’s like a family''}. P16 further shared \textit{``other apps are too popular and full of fake users i think people are better here and this platform boost my mood when i am bored''}.

\subsubsection{Graphological Identification}
\label{inkidentity}

\begin{figure*}[tb]
\centering
  \includegraphics[width=\textwidth]{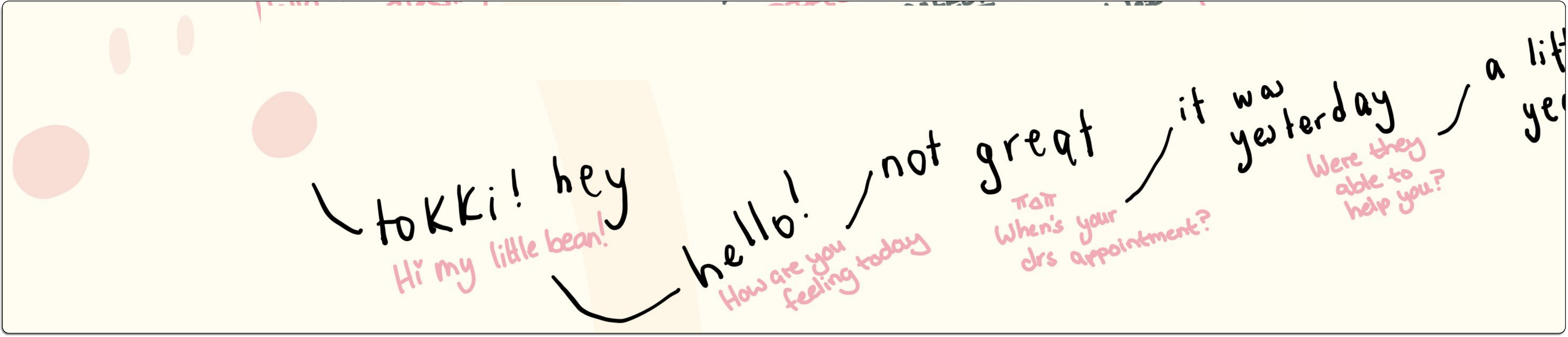}
  \caption{A supportive conversation between users on the GOS, where one user checks in on another's well-being and recent doctor's appointment. Notably, the user identifies the other through a familiar drawing.}
  \Description{A supportive conversation between users on the GOS, where one user checks in on another's well-being and recent doctor's appointment. Notably, the user identifies the other through a familiar drawing.}
  \label{fig:GOS_hand_writing}
\end{figure*}

\begin{figure}[tb]
  \centering
  \includegraphics[width=\columnwidth]{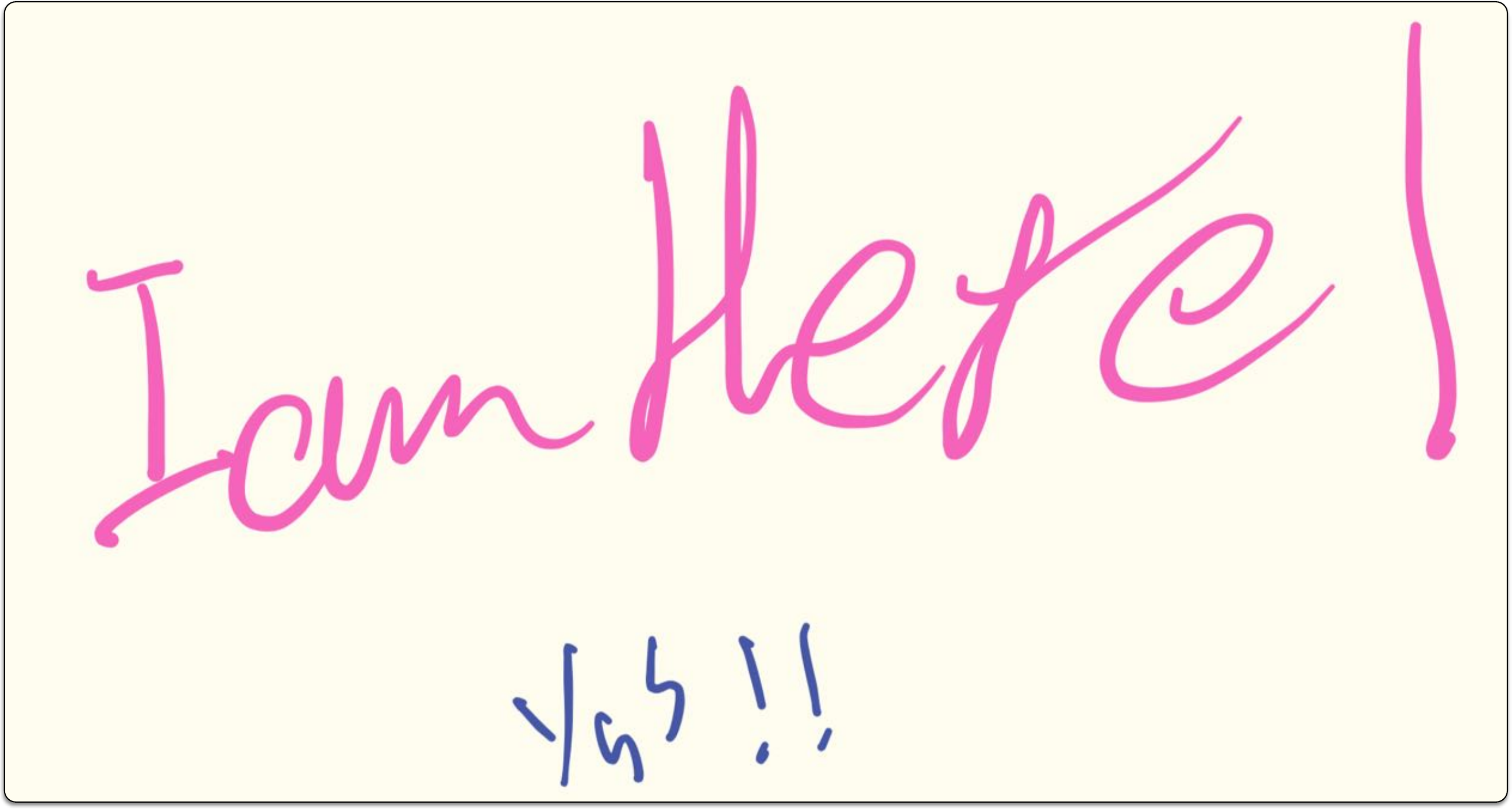}
  \caption{A user writes \textit{``I am here!''} in bold, pink handwriting, followed by a response of \textit{``Yas!!''} from another user. The second user's response suggests recognition, identifying the first user by their distinctive handwriting style.}
  \Description{Handwritten interaction where a user is recognized by their distinctive handwriting style.}
  \label{fig:GOS_next_handwriting}
\end{figure}

Our analysis of over 600 canvas pages on the GOS showed diverse patterns in handwriting and language use (see Table~\ref{tab:language_distribution}). While the majority of entries are in English (62.28\%), there is significant representation of Hindi (13.86\%) and Arabic (4.36\%), among other languages. Despite this linguistic variation, a prominent theme that emerges is the identification of users through their handwriting. Although the platform emphasizes anonymity, users often recognize one another based on handwriting style, ink color, or drawing patterns. In several instances, users appeared to identify prior collaborators solely through these visual cues. For example, in Figure~\ref{fig:GOS_hand_writing}, a user writing in black ink recognizes another user's pink drawing and greets them with \textit{"tokki! hey"}. In another instance (Figure~\ref{fig:GOS_next_handwriting}), a user with pink handwriting announces \textit{"I am here"}, and is immediately acknowledged with the response \textit{"Yas"}, confirming the recognition.

\begin{table}[tb]
  \centering
  \small
  \begin{tabular}{lrr}
    \toprule
    \textbf{Language} & \textbf{Count} & \textbf{Percentage} \\
    \midrule
    English & 629 & 62.28\% \\
    Arabic & 140 & 4.36\% \\
    Hinglish & 83 & 0.40\% \\
    Hindi & 44 & 13.86\% \\
    Chinese & 38 & 0.79\% \\
    Japanese & 20 & 1.98\% \\
    Turkish & 17 & 3.76\% \\
    Korean & 11 & 0.50\% \\
    Spanish & 8 & 8.22\% \\
    Italian & 5 & 0.20\% \\
    German & 4 & 1.68\% \\
    Russian & 2 & 1.09\% \\
    French & 2 & 0.20\% \\
    Filipino & 2 & 0.10\% \\
    Vietnamese & 1 & 0.10\% \\
    Thai & 1 & 0.10\% \\
    Portuguese & 1 & 0.10\% \\
    Hebrew & 1 & 0.20\% \\
    Bangla & 1 & 0.10\% \\
    \midrule
    \textit{Total} & 1010 & 100.00\% \\
    \bottomrule
  \end{tabular}
  \caption{Language usage distribution in GOS interactions.}
  \label{tab:language_distribution}
\end{table}

User interviews further confirmed this observation. Participants P02, P09, P15, and P17, P19 explicitly noted that digital handwriting has become a form of identification, allowing them to recognize their friends on the platform. P19 reported \textit{``by handwriting and emoji gestures I recognise my friends''}. This creates an interesting dynamic: while users appreciate the anonymity of the platform as seen by P03, who said, \textit{``Here, we can draw and collaborate without having to really know each other''}, they also develop a sense of familiarity through handwriting recognition. P05 contrasted this with Instagram, saying, \textit{``With Instagram, you know the name or origin of someone, but here, you don’t know anything about others''}. Over time, users' handwriting becomes recognizable, which serves as a creative signature that reflects their individual styles.

\subsubsection{Social Connections}
\label{socialconnections}

Another aspect frequently mentioned in the interviews is social connection. P02, P04, P09, P13, P16 and P19, and P20 expressed that they had formed close friendships on the platform, and these relationships were built not on verbal exchanges but on the unique creative expression using handwriting. P02 reported \textit{``I enjoy writing, specifically handwriting, to chat w/ users around the world. the connection feels more authentic. Also just love meeting + learning from others here hehe''}. Similarly, P09 wrote \textit{``I interact a lot with others + have lots of friends here''}. Many other users echoed similar experience; P13 said \textit{``I got a lot of safe friends here''}. 

\begin{figure}[tb]
\centering
  \includegraphics[width=0.4\textwidth]{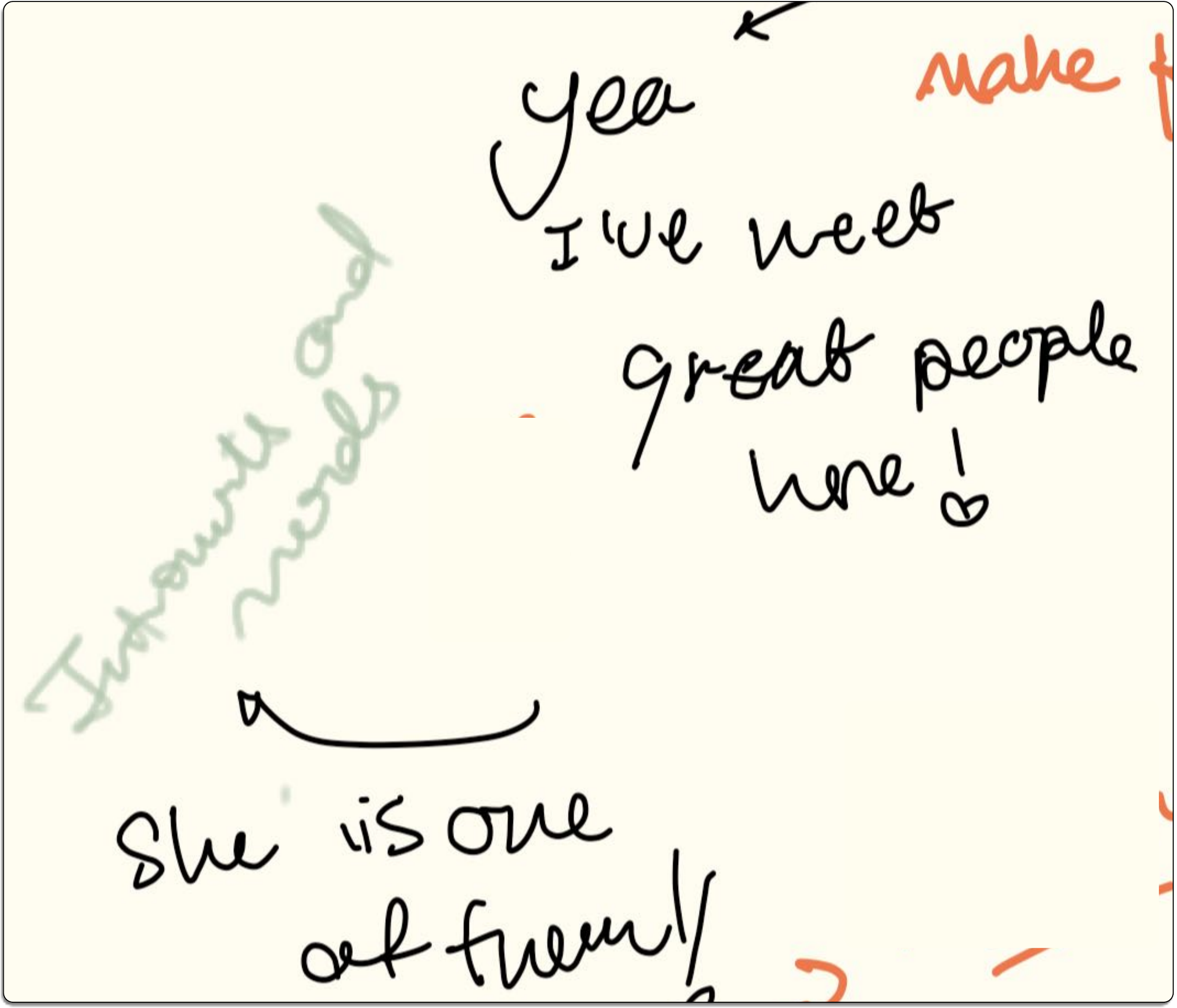}
  \caption{P20 emphasizes a connection by specifically pointing out the user with green handwriting.}
  \Description{P20 emphasizes a connection by specifically pointing out the user with green handwriting }
  \label{fig:GOS_social_connection}
\end{figure}

P19 emphasized their positive connections by drawing a heart and writing, \textit{``I have met great people here. Some have become really good friends, and we now talk frequently on Instagram''}. P20 in black handwriting shared a similar sentiment, stating, \textit{``I have met great people here''}, and specifically pointed to another user’s handwriting in green, noting, \textit{``She is one of them!!''} (Figure~\ref{fig:GOS_social_connection}).

\subsection{Conversation Analysis and Multimodal Discourse Analysis (RQ2)}
\label{sec:conversational}

Using CA and MDA, we analyzed real-time GIs across 10 GOS sessions (total 70 minutes), which uncovered visual and temporal dynamics of spontaneous exchanges among users. The analysis revealed a range of creative strategies that users on the GOS utilize to overcome social and contextual constraints while interacting. In this section, we summarize the CA results for turn-taking, sequence organization, and repair, and pair these with MDA findings on the intersemiotic functions of visual elements (termed \textit{visual cues}, \textit{spatial separation}, \textit{color coding}, \textit{canvas hopping}, and \textit{handwriting and paralinguistic cues}). Given that multi-modal resources are intricately tied to conversational structures on digital platforms, MDA complements CA by clarifying how these modes facilitate interaction. Therefore, we present both sets of findings together to highlight the multi-modal nature of communication on the GOS. For ease of reference, each user is assigned an arbitrary label (e.g., R1, B4) based on color or other visual cues. This does not correspond to their real identities. 

 \subsubsection{Turn-taking}
 \label{turntaking} Despite the seemingly `all over the place' appearance of the GOS, turn-taking mechanism is maintained smoothly and effectively by the users. They ensure a structured flow of their interaction through visual cues, spatial separation, color coding, canvas hopping, and handwriting and paralinguistic cues.

\begin{figure}[tb]
\centering
  \includegraphics[width=\columnwidth]{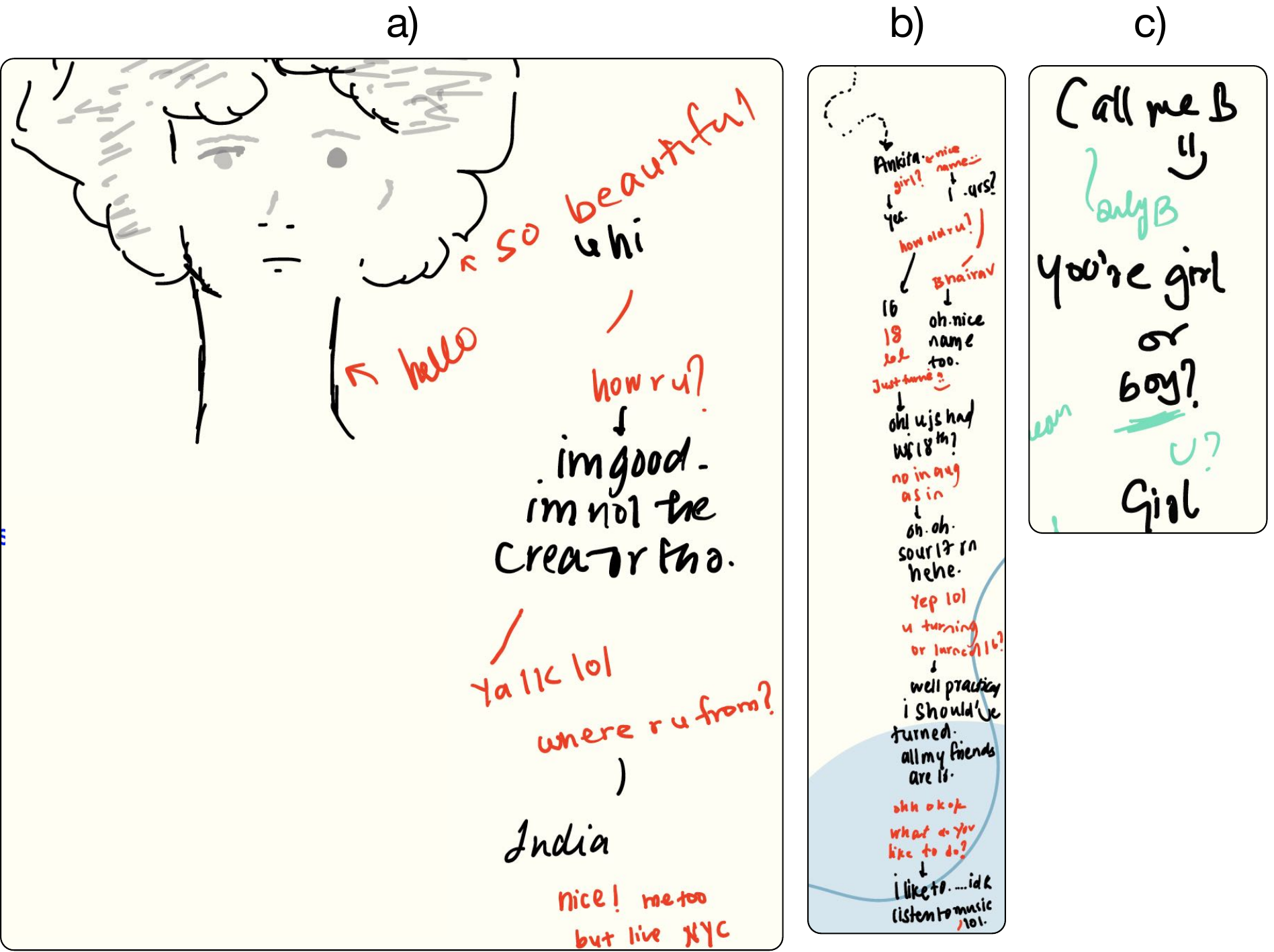}
  \caption{Examples of GI demonstrating creative turn-taking management through (a) natural conversation progression (b) visual connectors handling multiple utterances (c) non-verbal cues via underlining.}
  \Description{A vibrant collage of user-generated artwork from the GOS platform, featuring an eclectic mix of styles and forms}
  \label{fig:visualcues}
\end{figure}

\paragraph{\underline{\textbf{Visual cues}}} To elucidate this, consider two examples from the dataset. In Figure~\ref{fig:visualcues} (a), we observe two users: one writing in red (R1) and the other in black (B1). The conversation begins with R1 offering a compliment (\textit{``so beautiful''}) on the art and a greeting (\textit{``hello''}). B1 responds only to the greeting with ``hi'' and does not acknowledge the compliment, later clarifying that the artwork is not their creation (\textit{``im not the creator tho''}). Despite the initial ambiguity about the authorship of the artwork, the conversation flows naturally with both users building on each other’s responses.

Figure~\ref{fig:visualcues} (b), which is the continuation of Figure~\ref{fig:visualcues} (a), reveals a unique turn taking mechanism for managing multiple quasi-simultaneous utterances within the conversation. R1 presents two separate utterances: a compliment on the name (\textit{``nice name''}) and a question about gender (\textit{``girl?''}). B1 responds to both utterances using a visual cue such as arrows to link their responses directly to each specific prompt, effectively avoiding confusion, and ensuring that each utterance is addressed clearly. This approach differs from traditional text-based communication platforms, where all users can simultaneously post messages, with their order determined by when they are received by the server and the connection speed ~\cite{gonzalez2011conversation}. This linear approach of posting utterances often leads to overlapping and miscommunication~\cite{anderson2010turn}. 

Visual markers also play a significant role as non-verbal responses, contrasting to conventional text responses. For instance, in Figure~\ref{fig:visualcues} (c), when a user in black handwriting (B2) asks another user in light green handwriting (G2) if they were a boy or girl, G2 does not provide a direct answer. Instead, G2 uses a visual strategy by underlining the word ``boy'' in B2's question. These visual signals, combined with text, not only serve as responses but also function as focus markers to draw attention specifically to the part of the question relevant to the answer. This method demonstrates a form of unique communication resource within the GOS and reflects how users adapt to the platform's features to manage conversational turns.

\begin{figure}[tb]
\centering
  \includegraphics[scale=0.3]{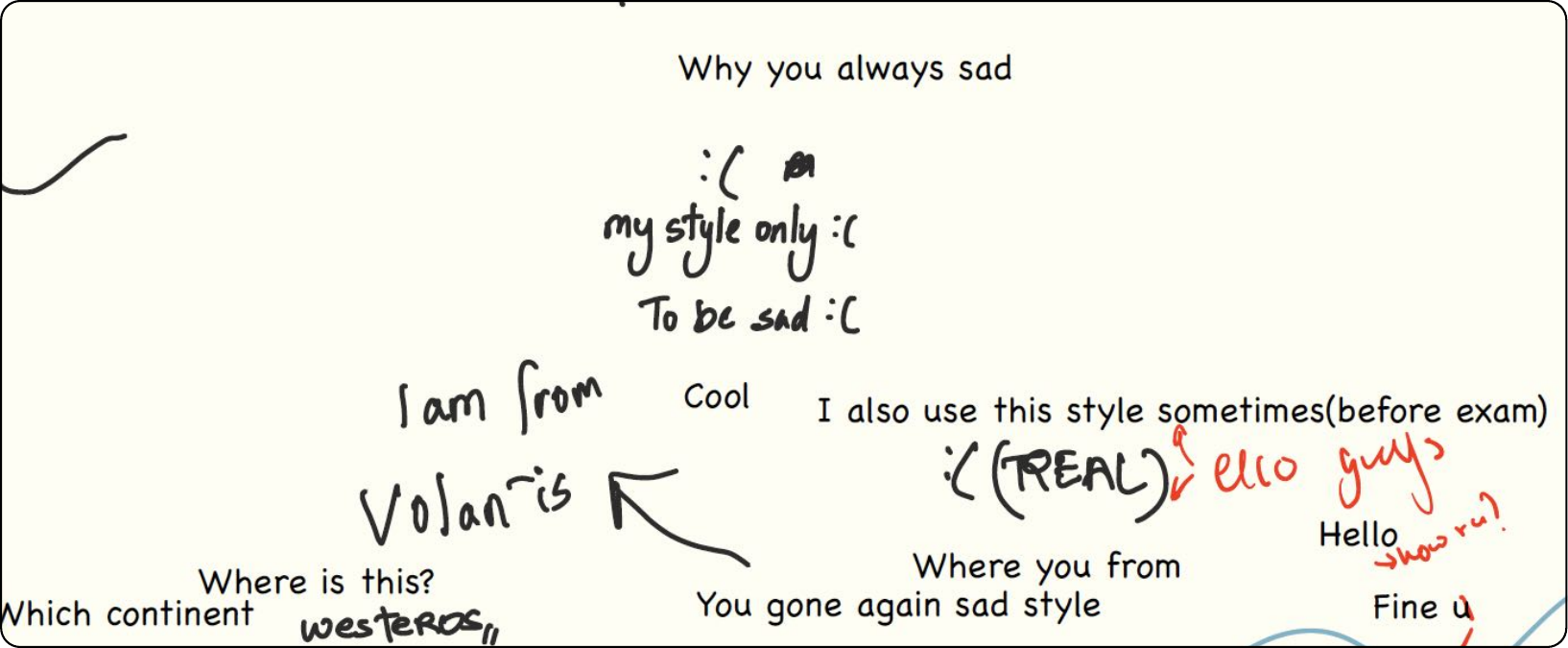}
  \caption{Managing multiple conversations with spatial separation.}
  \Description{Managing multiple conversations with spatial separation.}
  \label{fig:spatialseperation}
\end{figure}

\paragraph{\underline{\textbf{Spatial separation}}} We see another interesting turn-taking approach when more than two users are involved in a conversation. Figure~\ref{fig:spatialseperation} reveals that while engaging in simultaneous conversations, participants manage their interactions by utilizing different parts of the canvas. Here, the user (T) who types their messages, manages their conversation with two users: one using black handwriting (B3) and the other in red handwriting (R3). B3's conversation with T shifts to the left as soon as R3 jumps into the conversation with \textit{``Hello guys''}, prompting T to manage their interaction with R3 on the right side of the canvas. This strategic, non-linear positioning of text and spatial separation help distinguish between different threads of discussion and make the flow of overlapping conversations distinct. 

\paragraph{\underline{\textbf{Color coding}}} A recent study on visual note-taking highlighted the functional use of colors ``as a powerful way of bringing attention to content'' by highlighting key words and phrases ~\cite{zheng2021sketchnote}. However, on the current platform, this affordance extends beyond a cognitive tool to a more multifaceted role. Participants not only use different colors to create art forms but also identities. Color coding allows users to distinguish themselves and others in this multi-modal setting. This also helps identify each other's contributions without having to rely on context alone. 

\begin{figure*}[tb]
\centering
   \includegraphics[width=\textwidth]{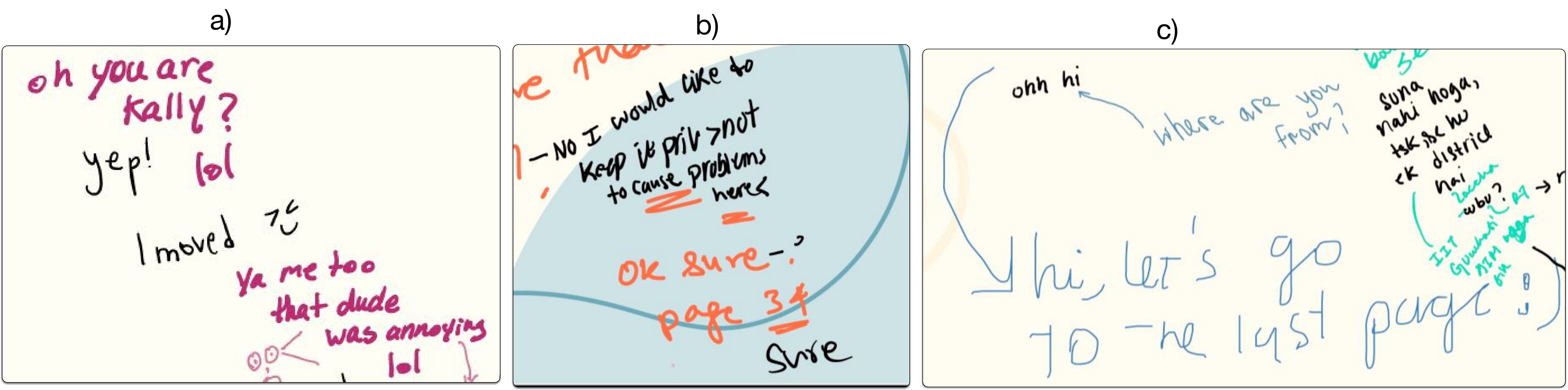}
  \caption{Canvas hopping. (a) Avoiding unwanted interactions (b) and (c) Minimizing visual clutter.}
  \Description{Canvas hopping. (a) Avoiding unwanted interactions (b) and (c) Minimizing visual clutter.}
  \label{fig:pagehopping}
\end{figure*}

\paragraph{\underline{\textbf{Canvas hopping}}} Another conversational tactic observed in GI included redirecting conversations through canvas hopping. Sometimes, participants involved in a conversation strategically switch to a different canvas page rather than continuing within the same thread to avoid conflict or uncomfortable exchanges. As the chat excerpt in Figure~\ref{fig:pagehopping} (a) illustrates, a user named Kally (likely a pseudonym, which users often develop over time as they form connections) in black handwriting and another user in pink handwriting navigate to a different canvas page to avoid an unwanted third party. They acknowledge their decision by comments like \textit{``I moved''} and \textit{``Ya me too that dude was annoying lol''}, and seamlessly restart their conversation. This highlights both platform's flexibility and users' preference for facilitating a more controlled and positive dialogue environment.

In addition, canvas hopping is also seen when the canvas becomes overly busy with drawings and interactions. Users switch to a new page to mitigate visual clutter and continue their conversation in a less congested space. For instance, the chat excerpts in Figure~\ref{fig:pagehopping} (b) and Figure~\ref{fig:pagehopping} (c) show users expressing a preference to move with \textit{``page 34''} and \textit{``hi, let's go to the last page''} to avoid disturbances. This strategy is akin to face-to-face conversations where people might choose to relocate to a different part of a crowded room to create a comfortable conversational space.

\begin{figure*}[tb]
\centering
  \includegraphics[width=\textwidth]{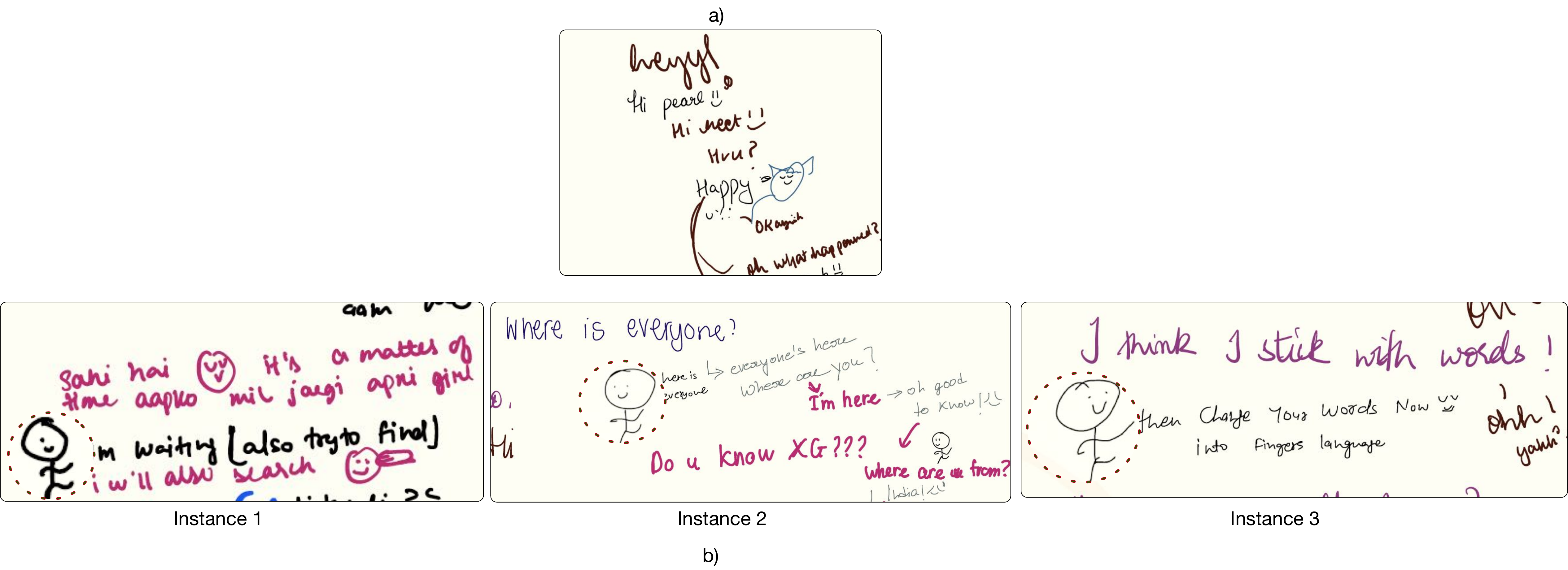}
  \caption{Handwriting and paralinguistic cues as identity markers. (a) Typical use of exclamation mark in greetings by a user (b) A user's distinctive stick figure art for recognition. (Note: circled for clarity)}
  \Description{Handwriting and paralinguistic cues as identity markers. (a) Typical use of exclamation mark in greetings by a user (b) A user's distinctive stick figure art as an identity marker.}
  \label{fig:paralinguistic}
\end{figure*}

\paragraph{\underline{\textbf{Handwriting and paralinguistic cues}}} Interestingly, despite anonymity, participants who are regular on the GOS recognize each other by their unique handwriting styles and paralinguistic cues such as specific drawings or distinctive patterns. This finding was also corroborated in semi-structured interviews (refer to ~\ref{inkidentity}). For instance, in Figure~\ref{fig:paralinguistic} (a), the user identified as Reet, in black handwriting, recognizes Pearl in brown handwriting by noting the distinctive use of an exclamation mark in Pearl's greeting (\textit{``heyy!''}), which is consistently included in their other interactions (not all shown here). Some users draw distinct art styles along with their text that help other users recognize them. Figure~\ref{fig:paralinguistic} (b) highlights three separate instances, each from different conversations on different days, where a user (in black handwriting) consistently draws a specific stick figure alongside their text to signal their identity and convey a friendly tone. 

\subsubsection{Sequence Organisation} The analysis showed that the organization of conversational sequences gets more varied and complex in GI depending upon the number of participants.

\begin{figure}[tb]
\centering
  \includegraphics[width=\columnwidth]{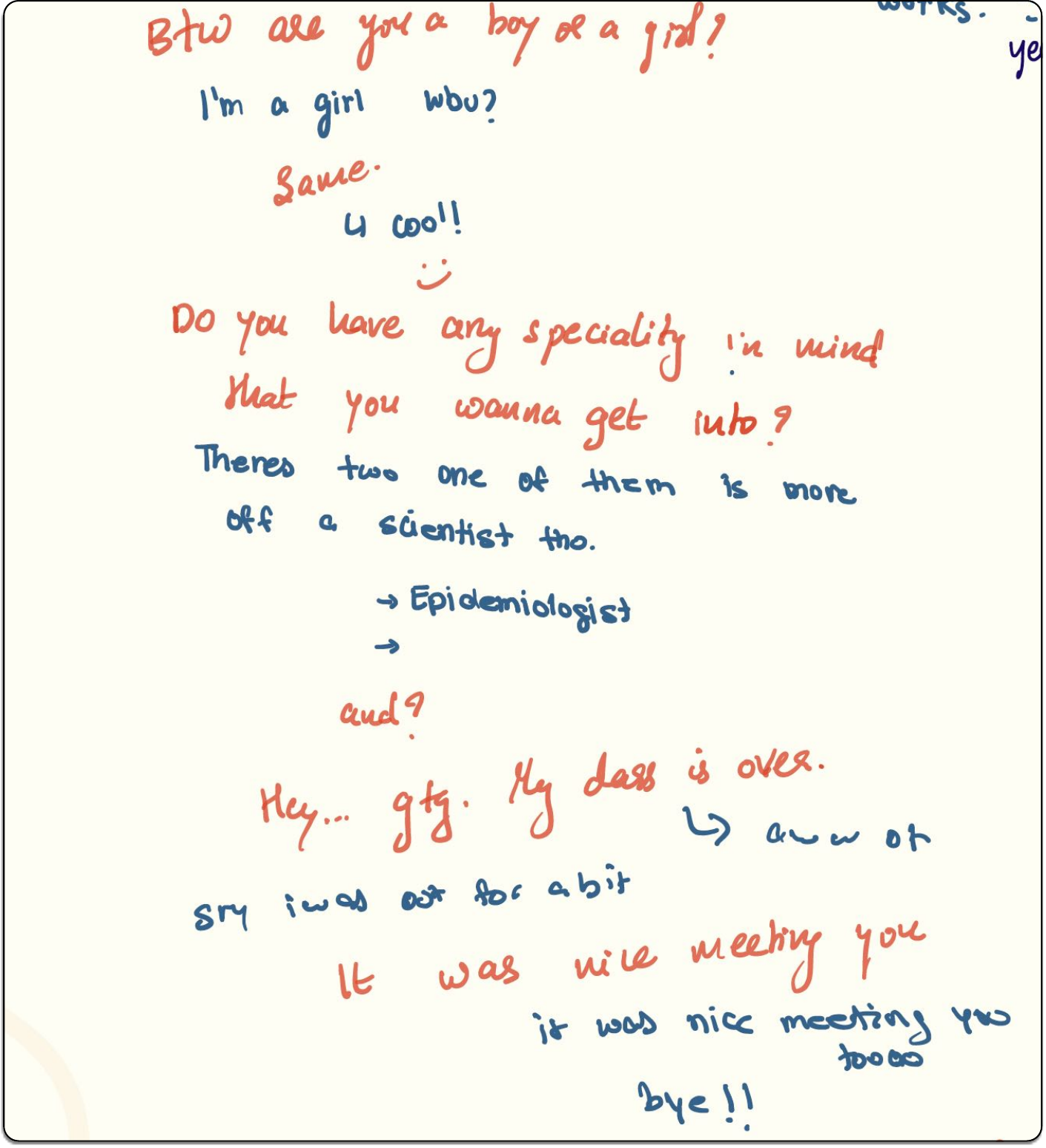}
  \caption{Sequence organisation in a one-to-one interaction using adjacency pairs.}
  \Description{Sequence organisation in a one-to-one interaction using adjacency pairs}
  \label{fig:onetoone}
\end{figure}

\paragraph{\underline{\textbf{One-to-one interaction}}} In one-to-one interactions, sequences are logically expanded through clear adjacency pairs. Figure~\ref{fig:onetoone} depicts the use of several adjacency pairs of questions and answers followed by a brief exchange of confirmation (\textit{``same''}) and acknowledgment (\textit{``cool''}, \textit{:)} ). A post-expansion phenomenon is also observed when a user in red handwriting (R4) prompts their co-participant in blue handwriting (B4) for more information with \textit{``and?''}. Post-expansion refers to the statements or sequences that follow the main action (in this case: \textit{``Do you have any speciality in mind that you wanna get into?''}) in order to provide clarification, extend the ongoing interaction or offer extra information~\cite{stivers201310}. On not receiving any response from B4 (likely because B4 is not present), R4 follows up by \textit{``Hey..gtg. My class is over''}, which guides the conversation towards a close with B4 acknowledging it (\textit{``aww ok''}) and using a post-expansion (\textit{``sry i was out for a bit''}) to reinforce a friendly tone.

\begin{figure}[tb]
\centering
  \includegraphics[width=0.8\columnwidth]{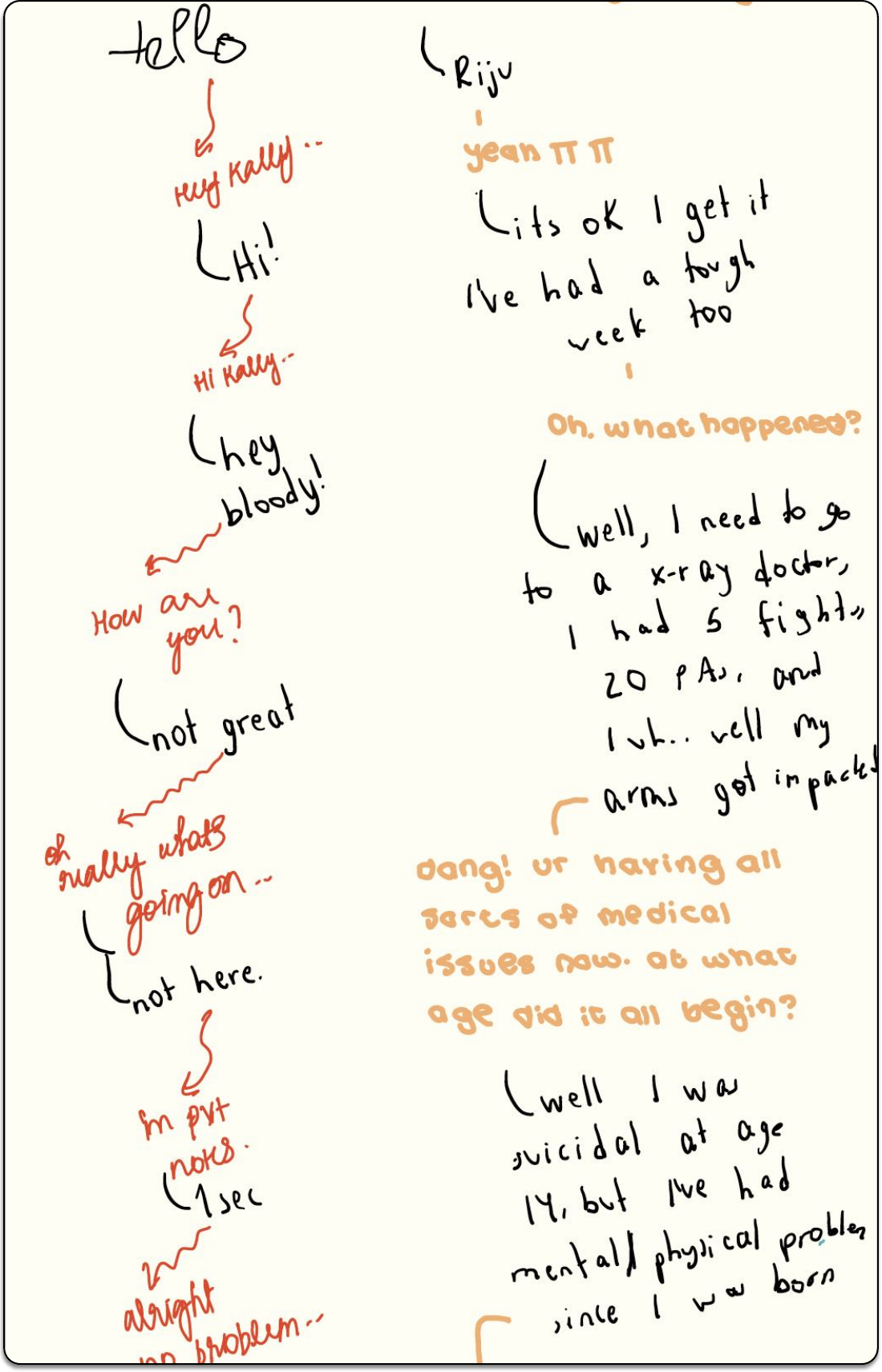}
  \caption{Interactions among three users illustrating emotional engagement in sensitive contexts.}
  \Description{Interactions among three users illustrating emotional engagement in sensitive contexts.}
  \label{fig:onetothree}
\end{figure}

 \paragraph{\underline{\textbf{Interaction among three users}}} A similar pattern emerges in interactions involving groups of three. Figure~\ref{fig:onetothree} illustrates such a conversation, where a person identified as Kally in black handwriting is engaged in conversation with two other users O5 (in orange handwriting) and R5 (in red handwriting). Both interactions demonstrate a careful organization of adjacency pairs where each response aligns with the emotional tone set by the previous message, for instance, when R5's question \textit{``How  are you ?''} is met with a response \textit{``not great''}. This response highlights the principle of conditional relevance, as it not only signals the respondent's current state but also implies that there is something more to discuss. The response opens the conversational space for the follow up question (\textit{``oh really whats going on..''}), which directs the conversation towards delving deeper into the respondent's emotional state.
 The responses \textit{``not here}, \textit{1 sec''} indicate that the user is not comfortable sharing details in the current setting. The follow-up reply \textit{``alright no problem''} demonstrates acceptance and understanding and gracefully brings their conversation to a close. Similarly, Kally and O5 while discussing sensitive topics such as suicide and medical issues, maintain a delicate balance between showing empathy through expressions like \textit{``oh what happened''}, and \textit{``dang''}. O5 prompts further inquiry with a question like \textit{``at what age did it all begin ?''}, without overwhelming the other participant.

\begin{figure}[tb]
\centering
  \includegraphics[width=\columnwidth]{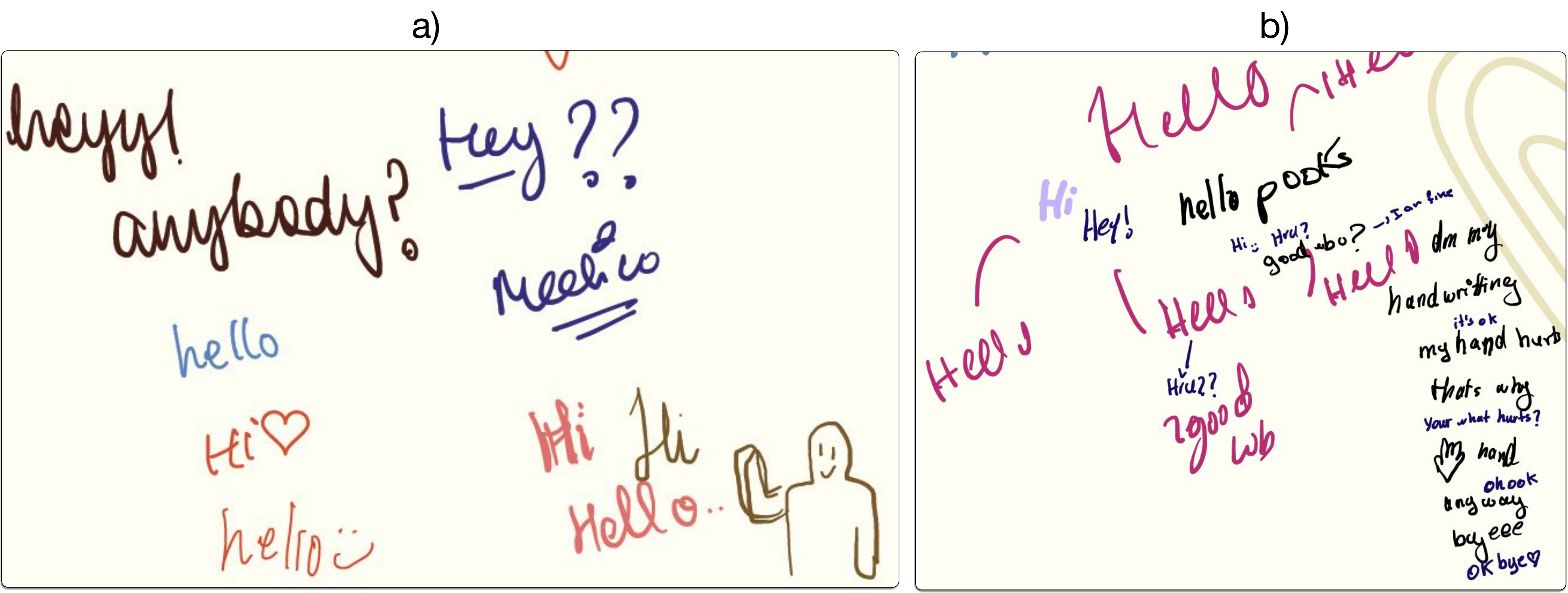}
  \caption{Complexities in multi-user interactions. (a) Brief exchanges (b) Conversation fragmentation due to selective engagement.}
  \Description{Complexities in multi-user interactions. (a) Brief exchanges (b) Conversation fragmentation due to selective engagement.}
  \label{fig:morethanthree}
\end{figure}

  \paragraph{\underline{\textbf{Interaction among more than three users}}} While most GI involved two or three users, we also observed group conversations involving more than three users in a conversation. Two notable patterns tend to emerge in these group interactions. First, an interaction is often restricted to brief exchanges such as simple greetings as seen in Figure~\ref{fig:morethanthree} (a). Second, when a conversation extends beyond initial adjacency pairs, usually two participants are actively involved in continuing the dialogue and driving the conversation forward. These patterns are also commonly observed in conventional chatrooms with multiple overlapping conversations. However, while conventional chatrooms often witness this phenomenon due to imposed linearity~\cite{herring1999interactional} and the platform's  control over turn-taking ~\cite{al2023pragmatic}, GOS provides users with more autonomy. Here, users have greater control over managing conversations through visual and spatial tools, which allow them to manage multiple simultaneous conversations. Consequently, fragmented conversations on the GOS are often due to factors like the absence of responses from other participants, lack of interest, or delayed entry of a user into an ongoing conversation, as seen in Figure~\ref{fig:morethanthree} (b). In this example, four participants, identified by handwriting in pink (P6), violet (V6), blue (B6) and black (BL6), engage in a conversation. P6 initiates the conversation with a \textit{``Hello''} to which the other users respond with similar greetings. Interestingly, P6 maintains engagement with all participants by replying to each greeting again by another  \textit{``Hello''}, using connecting lines. B6 expands the conversation by asking both P6 and BL6, \textit{``Hru?''}. While both P6 and BL6 respond with \textit{``good wbu?''}, B6 chooses to continue the conversation only with BL6, leaving P6's response unanswered. This results in a more well-defined conversation between B6 and BL6.

\subsubsection{Repair}
\label{repair}
Although not frequent, the repair phenomenon plays an essential role in conversation dynamics on the platform. Our analysis showed that fewer instances of repair may be attributed to the focused attention required in the writing process, which is inherently unimanual in nature ~\cite{lee2020smartphone}. Writing represents a more natural input method compared to digital tapping/typing, where errors are typically caused due to small screen sizes, inaccurate finger touch~\cite{siek2005fat,holzinger2012answer} and autocorrect failures ~\cite{alharbi2022text}. However, we did observe instances of repair, although they were differently expressed due to both the platform's affordances and the mode of communication i.e., writing on iPad using a stylus.

\begin{figure}[tb]
\centering
  \includegraphics[width=\columnwidth]{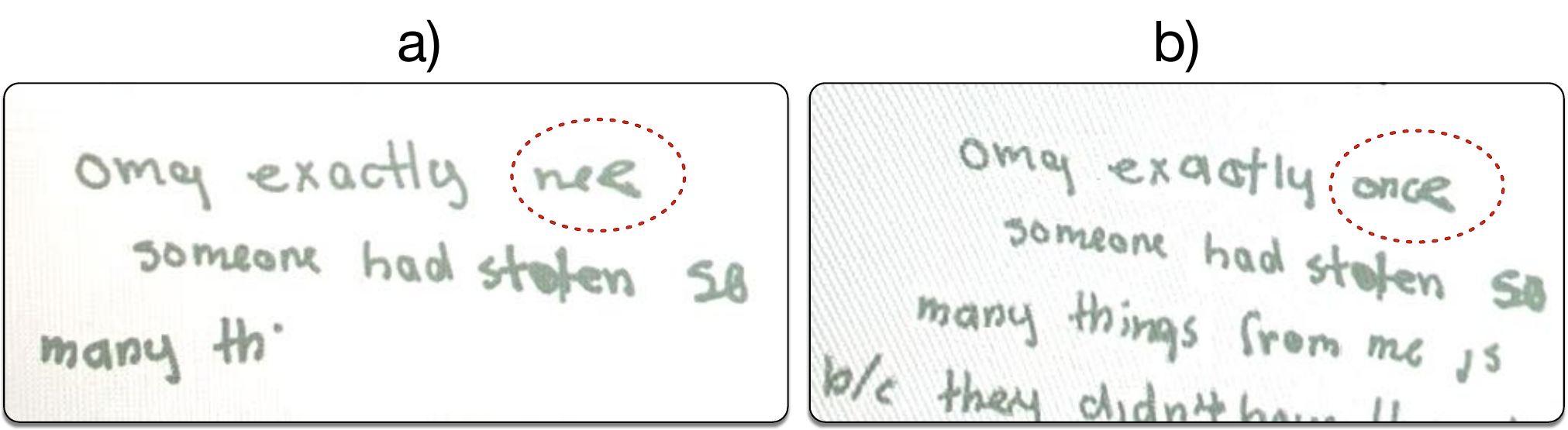}
  \caption{Self-repair in real-time. (a) Initial message with a typo ~\textit{"nee"} (b) Corrected message ~\textit{"once"} showing real-time self-repair (Note: Highlighted for clarity).}
  \Description{Self-repair in real-time. (a) Initial message with a typo (b) Corrected message showing real-time self-repair ~\textit{"once"} (Note: Circles are added for clarity).}
  \label{fig:selfrepair}
\end{figure}

\paragraph{\underline{\textbf{Self-repair}}} A clear preference for self-repair is seen in GI, which became evident through repeated instances across different conversations where users visibly corrected their writing mid-construction.
This phenomenon is akin to other chat platforms. However, while on platforms such as Facebook chat, ``participants are able to perform such repairs during message construction, so that the recipient is unaware of it'' or after the message has been sent~\cite{meredith2014repair}, the GOS allows users to monitor each other's writing in real-time, including any self-repairs made during message construction. For instance, in Figure~\ref{fig:selfrepair} (a), a user in green handwriting initially wrote \textit{``nee''}. Although the handwriting is somewhat ambiguous, it appears they intended to write \textit{``once''}, which was subsequently corrected, as shown in Figure~\ref{fig:selfrepair} (b). 

\begin{figure}[htb]
  \centering
  \includegraphics[width=\columnwidth]{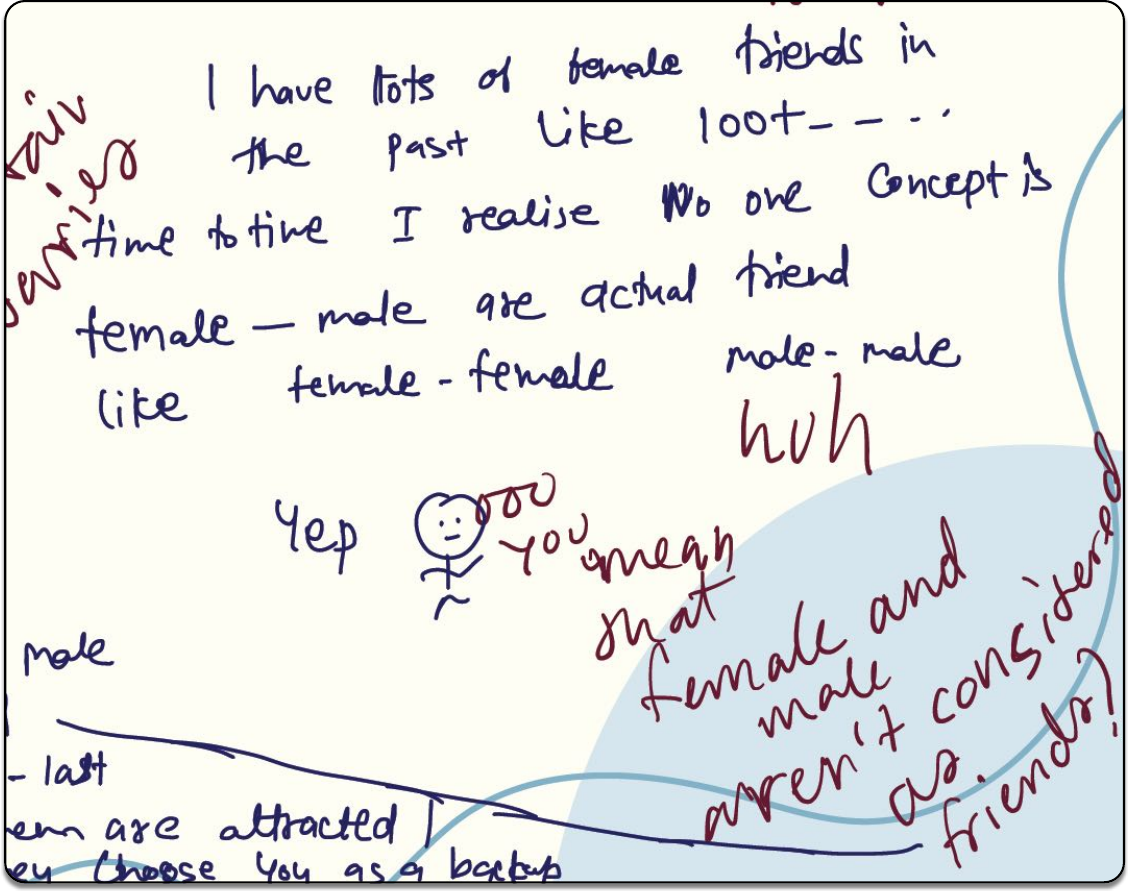}
  \caption{Other-initiated repair. R7 requests clarification for better understanding.}
  \label{fig:otherrepair}
\end{figure}

\paragraph{\underline{\textbf{Other-initiated repair}}} The other-initiated repairs in GI are carried out mainly to clarify the meaning behind the initial speaker's original message as illustrated in Figure~\ref{fig:otherrepair}. In the excerpt, the users are discussing the nature of friendship. 

The user in red handwriting (R7) seems to be confused at the statement of friendship made by the initial user in blue handwriting (BL7). R7 responds with ~\textit{``huh''} to indicate lack of understanding of the idea presented by BL7. R7 further elaborates by asking ~\textit{``ooo you mean''}, which is an instance of clarification request (other-initiated repair) aimed at repairing the misunderstanding. Here, the repair sequence involves both the clarification request and the subsequent clarification (~\textit{``that female and male aren't considered as friends?''}) by R7. This sequence helped ensure that both participants had a shared understanding of the topic and avoided any potential misunderstandings. 

\begin{figure}[tb]
\centering
  \includegraphics[scale=0.35]{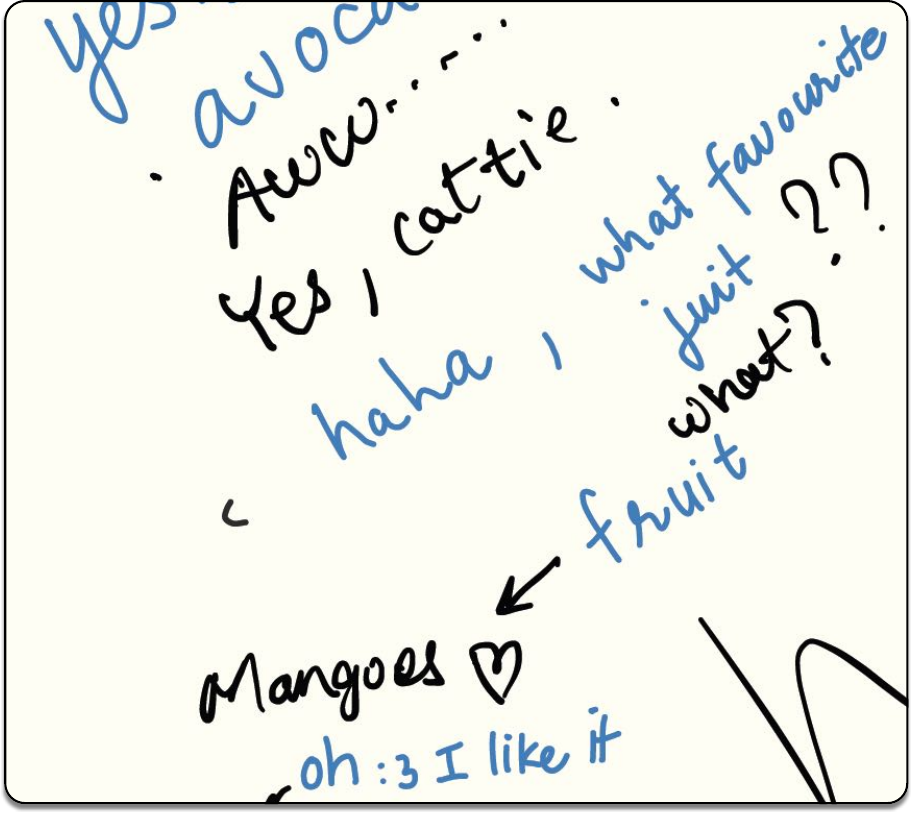}
  \caption{Stylus induced misspelling and correction. ~\textit{Juit} corrected to ~\textit{fruit} following clarification request.}
  \Description{Stylus induced misspelling and correction. ~\textit{Juit} corrected to ~\textit{fruit} following clarification request.}
  \label{fig:implicitrepair}
\end{figure}

\paragraph{\underline{\textbf{Implicit repair}}} Additionally, we observed that not all misspelled words are corrected as long as the intended meaning is clear from context. This may be due to the perception of online interactions as informal, where non-standard spellings are generally acceptable among users ~\cite{meredith2014repair}. However, the misspelled words that users seek further clarification on emerge specifically due to the writing medium -- stylus. Despite improved accuracy, digital styluses can occasionally introduce imprecise movements on the screen while writing ~\cite{fernandez2020live, annett2020pen}, prompting participants to initiate repair as seen in Figure~\ref{fig:implicitrepair}.
In this particular instance, the misspelling \textit{``juit''} by the user in blue handwriting (BL8) is influenced by the stylus use, as it is evident when BL8 later corrects it to \textit{``fruit''} after the user in black handwriting asks for clarification with \textit{``??''} and \textit{``what?''}. 

\section{Discussion}
\label{sec:discussion}

In this paper, we explored social engagement and creativity within the novel environment of Graphonymous Online Space (GOS), along with the conversational dynamics emerging from a new form of interaction we term Graphonymous Interaction (GI). ~\add{In this section, we first unpack the specific GOS features, design affordances, and user practices that enable creativity, emotional expression, and social connection}. In Section ~\ref{applications} we discuss our findings by highlighting the potential applications of GOS and GI, while also addressing the challenges and risks associated with this emerging mode of interaction. We conclude by acknowledging the limitations of our study and outlining open questions that offer promising directions for future research.

\subsection{\add{Features of GOS that fosters Social Engagement and Creativity}}  

\add{The diversity of artistic styles observed in our dataset (see Table~\ref{tab:drawing_styles}) which consisted of 11 distinct styles and 724 total drawings reflects how specific GOS features enable creative participation across skill levels. Three main features we identified on GOS worked in tandem: (1) the absence of artistic gatekeeping mechanisms, (2) pseudo-anonymity through handwriting, and (3) an hourly reset. The lack of formal filters and quality expectations from the community (unlike platforms such as FlockMod, where room owners can restrict access via passwords~\cite{FlockMod2024}) allows stick figures and watercolors to coexist equally. Pseudo-anonymity, where users remain unnamed yet recognizable through handwriting, creates psychological safety. We observed that participants shared vulnerable emotions (Figure~\ref{fig:GOS_emotions}) precisely because they felt visually recognizable yet protected from persistent reputation consequences. The hourly reset further reduces judgment risk by eliminating permanence; users treat each session as a ``clean slate'' for low-stakes experimentation. }  \add{Similarly, social engagement (evidenced in social connections theme~\ref{socialconnections} and interview  quotes about ``family'') emerges from three other affordances: (1) spatial, non-linear interface design, (2) real-time visual awareness, and (3) informal group formation mechanisms. The spatial canvas eliminates the sequential constraint of text-based chat, allowing users to exploit spatial separation, color coding, and visual markers to manage multiple simultaneous conversations (Section~\ref{turntaking}). Real-time visual awareness i.e., users monitoring each other's stylus strokes in real time, reduces repair episodes (Section ~\ref{repair}) compared to messaging systems with lag-based temporal distortion. Informal group formation (see Figure~\ref{fig:GOS_intellectualengagement}) removes hierarchical barriers, allowing egalitarian collaboration even among strangers, supporting co-play (distinct from competition-driven gaming or metrics-driven social platforms). }

\add{Intellectual engagement games and problem-solving documented in Section~\ref{intellectengagement} specifically emerges from the multi-modal affordance of combining drawing and writing. Drawing a math graph, or collaborative puzzle seems natural on GOS but cumbersome in text-only systems. This likely contributes to understanding why 47 games were started and mathematical discussions arose (see Table~\ref{tab:subject_distribution}), even though the platform did not include any formal ``game'' structures.}

\subsection{Potential Applications of GOS and GI}
\label{applications}
\subsubsection{GOS as a Social Medium for Democratizing Art}


Our findings suggest that GOS provides a highly accessible medium for visual expression, which helps to reduce barriers to artistic participation. We observed a wide range of drawing styles, from intricate manga and silhouettes to simple stick figures and kawaii inspired illustrations, all reflecting an inclusive aesthetic culture. Unlike other creative platforms that value professionalization, GOS fosters a non-judgmental environment where both novice and experienced artists can contribute, co-create, and be seen. Crucially, feedback within the GOS appeared highly personalized. Users offered comments using varied ink colors, handwritten notes, and idiosyncratic rating schemes (e.g., verbal encouragement, hearts, or numeric scores). This suggests a form of intimate peer critique rarely seen on mainstream platforms. Unlike dominant image-sharing networks, where feedback mechanisms are flattened into likes or algorithmically filtered comments~\cite{exampleuser2024reddit}, GOS fosters a more human-centered and dialogic mode of creative interaction.

During our study, the intensification of the Israel-Palestine conflict offered a glimpse into how GOS might also function as a site for visual activism. Rather than sharing links or hashtags, users expressed solidarity through deliberate drawings such as the Palestinian flag into their artworks or surrounding pieces with symbolic frames and hearts. These contributions, while small and subtle, point to the emergence of what we might call \textit{\textbf{graphonymous activism}}: visual, anonymous, non-verbal gestures of political expression that are embedded within GOS. This mode of engagement departs from conventional social media activism ~\cite{ricardaactivism} by abstaining from social media metrics in favor of quieter, more contextual forms of solidarity. These observations, taken together, suggest that GOS has the potential to be more than just a creative outlet. It could also serve as a platform for fostering micro-communities of mutual support, artistic growth, and visual activism. 

\subsubsection{GOS as a Social Medium for Mental Health}

The study of emotional sharing in social media has emerged as a significant research area within HCI~\cite{bazarova2015social}. In our findings, users on the GOS demonstrated an ability to express emotions through digital art and handwriting. Notably, several participants described the platform as a space for emotional expression and social support. Users shared personal experiences and emotions more openly than they might on conventional platforms, often describing the GOS as "\textit{safe}" and "\textit{supportive}". Many examples of users seeking help on the GOS were found during the thematic analysis (see sec~\ref{sharingandsupportongos}), ``\textit{Can someone talk to me?}'' ``\textit{I’m feeling depressed}'', "\textit{I'm having a panic attack}", ``\textit{I have issues in my work can someone help me}''. Another important aspect that emerged from the data is the subtle signaling of emotional or psychological needs on the GOS. For instance, a user used a shorthand phrase such as ``PA = Panic...Attack'' (Figure~\ref{fig:GOS_support_pain}). What is notable is that these signals were not ignored but responded to, either by starting a conversation or offering support. It can therefore be assumed that GOS fosters a quiet attentiveness, where emotional cues are acknowledged in a non-intrusive, low-pressure way. 

The platform's design further shapes this environment through its anonymous structure and hourly reset. Because the canvas refreshes every hour, users' posts disappear, thereby reducing the risk of being scrutinized or trolled over time. These dynamics stand in contrast to contemporary mainstream social media, where personal sharing is often constrained by identity performance, surveillance, and commercial incentives ~\cite{wilson2020antisocial}. ~\citet{onlypositiveemotions} corroborates this finding in their qualitative study, which shows that the primary usage of Snapchat, Facebook, and Instagram is for sharing positive emotions. Negative emotions are typically shared in private messages, with users exercising caution before posting, as they may face social repercussions~\cite{onlypositiveemotions,bazarova2015social}. 

Based on our findings, we posit that employing handwriting and drawing as means of conveying negative emotions within a social space can be significantly more effective compared to traditional text-based messages and photos, as informed by our interview participants. These latter modalities are generally regarded as less effective in conveying negative emotions within social media space, and are at risk of being misinterpreted ~\cite{onlypositiveemotions}. Additionally, the ability to express emotions in ones own handwriting and drawing aligns with prior work on the therapeutic potential of artistic expression ~\cite{drake2013children,drakedrawing,yan2021regulatory}. This suggests that GOS may facilitate a form of ambient peer support for mental health and self-regulation through creativity. GOS offers an alternative model: one where anonymity, visual expression, and minimal platform instrumentation create space for emotionally resonant interactions. 

\subsubsection{GOS as a Social Medium for Community Building}

Based on our findings, the GOS supported forms of collective engagement that blended play, problem-solving, and co-creation. This aligns with the longstanding interest of Computer Supported Cooperative Work (CSCW), which aims to research and develop technologies that facilitate collaboration and partnerships among individuals, teams, and communities~\cite{10.1145/3359250}. The group building activities on the GOS offer an initial glimpse into its pivotal role in supporting a collaborative and dynamic community. From collaborative art work to strategic games and problem solving, these user-led interactions show how GOS allows users to shape their own shared environment. More precisely, the formation of groups often varies depending on the objective of the activity. 

On the GOS, we found users invite others to join an activity by initiating it in an egalitarian and non-hierarchical fashion.~\add{This collective engagement is actively enabled by GOS's collaborative affordances.} For instance, a user draws a Tic-Tac-Toe grid, makes the first move and then waits for someone to join by writing ``\textit{Anyone?}''. This gives agency to others, even without any prior relationship and warmly welcomes anyone willing to play. This form of group formation differs from computational systems that automate group formation processes by matching users and tasks, resulting in a mechanical process lacking motivation ~\cite{10.1145/1871437.1871515,7445184}. The dynamics on the GOS, on the other hand, reflect what Boutillier et al.~\cite{boutillier2020collaborative} describe as collaborative atmospheres that foster creativity and interdisciplinary interaction. 

On GOS, this atmosphere emerges organically, not through system-level orchestration, but from user practices: responding kindly, adding to drawings, playing together. We propose that the conventional understanding that individuals are more inclined to collaborate with those who share similar characteristics ~\cite{bardzell2008blissfully} requires reconsideration in the context of GI. On the GOS, users are often unfamiliar with those who join their activity, yet collaboration takes place without any conflict (see Figure~\ref{fig:GOS_art_math}). These moments are characterized by mutual responsiveness and fluid turn-taking, as users collaborated on content in a non-hierarchical manner. In an online space like GOS, we see a renewed form of social co-play that is less about competition or metrics, and more about shared presence, synchronous participation, and expressive collaboration. 

\subsubsection{GOS as Social Media: Expanding the Form-From Design Space}
\label{GOSinfromfrom}
In a recent 2024 CSCW research paper,~\citet{10.1145/3641006} introduced a two-dimensional framework for analyzing social media platforms. This framework categorizes platforms based on their interaction structure (``\textit{Form}'': flat or threaded) and content sourcing (``\textit{From}'': spaces, networks, or commons). This framework helps us understand how different social systems facilitate various forms of discourse, engagement, and governance. Within this conceptual landscape, GOS represents an interesting and relatively unexplored configuration, ``\textit{Flat-Space}'' environment that breaks away from conventional text and feed-based platforms.

GOS's ``\textit{flat}'' nature comes from its lack of algorithmic hierarchies, threaded responses, or ranked timelines. Instead, users interact in real-time on shared canvases, contributing handwritten content and drawings without any system enforced reply structures or visibility algorithms. It functions as a ``\textit{space}'' because interactions occur within discrete, bounded canvases (similar to chat rooms or collaborative documents) rather than personalized timelines or global feeds. What makes GOS particularly distinctive is how it extends the Flat-Space model through stylus-based, visual, and graphically embodied communication. Unlike traditional Flat-Space platforms such as Slack or early IRC with their purely textual interfaces, GOS enables spatially-oriented, multi-modal conversations built around handwriting, sketching, and visual annotation.

This departure from standard media formats expands the Form-From model both theoretically and empirically. It challenges the assumption that social media ``\textit{Form}'' is primarily defined by text organization (threaded vs. linear), by demonstrating how visual composition, spatial elements, and handwriting-based authorship create their own logic of turn-taking, visibility, and conversation flow. Our analyses reveal that users employ visual cues such as arrows, ink colors, positioning, and drawing layers to indicate conversational progression. These practices serve similar functions to threading or quoting in text interfaces, but with distinct characteristics.

Additionally, GOS complicates typical relationships between anonymity and identity that are commonly present in Flat-Spaces. Despite being nominally anonymous, participants frequently recognize each other through handwriting styles, a phenomenon we have termed ``\textit{graphological identification}''. This creates a form of ambient identity that is neither fully persistent nor completely disposable, which raises interesting questions about authenticity and vulnerability in anonymous creative platforms. Unlike platforms such as Whisper (Flat-Commons)~\cite{wang2014whispers} or BeReal (Threaded-Network)~\cite{10.1145/3613904.3642690}, GOS enables personal expression without encouraging performative content creation or algorithmic exposure. This aligns with recent calls for alternatives to networked, metrics-driven social systems.

By occupying this unique position in the Form-From framework, GOS shows that alternative social media can support emotional, collaborative, and expressive interaction without depending on profiles, feeds, or threaded conversations. In doing so, it not only exemplifies a new category of Flat-Space systems but also encourages broader thinking about how social interaction might be structured around visual, ephemeral, and ambient forms of participation. 

\subsection{Challenges and Risks}
\subsubsection{Accessibility Challenges of Graphonymous Online Spaces}
\label{sssec:accessibility}
GOS and GI introduce significant accessibility barriers. In their current form, these platforms risk excluding users with visual and motor impairments due to their reliance on freeform, handwriting based interaction without structured assistive features. Without deliberate design interventions, GI may replicate patterns of exclusion common in creative and visually dominant digital spaces. For users with motor impairments, generative AI systems could offer meaningful support by converting typed input into customized digital handwriting. This would not only allow seamless participation in GI but could also preserve the expressive and identifiable quality of handwriting, which plays a key social role in GOS. For users with visual impairments, prior work has shown that real-time support through crowd sourcing ~\cite{10.1145/1866029.1866080} and visual question answering systems ~\cite{Gurari_2018_CVPR} can bridge accessibility gaps. However, these approaches must be extended to meet the unique demands of GOS. 
As our analysis shows, communication in GI relies on spatial positioning, visual cues, and implicit turn-taking mechanisms such as canvas hopping, color coding, and directional markings, screen readers and accessibility tools must be adapted to recognize and interpret these practices, not merely describe images. Future accessibility systems could benefit from incorporating recent advances in handwriting recognition ~\cite{graves2009offline,graves2009novel} and identification ~\cite{rehman2019automatic,alkawaz2020handwriting,javidi2020deep}, to support non-visual forms of authorship and identification in graphonymous environments. Accessibility in GI and GOS requires more than retrofitting existing tools. It requires a new understanding of interactional accessibility in visually appealing, low-structured digital environments. 

\subsubsection{Pseudo-Anonymity and False Security}
While GOS presents itself as an anonymous and identity-agnostic platform, our findings reveal a more complex reality. Several participants noted that handwriting functioned as a kind of personal signature as discussed in section ~\ref{GOSinfromfrom}, allowing them to recognize past collaborators or even describe their handwriting as a form of digital identity - \textbf{\textit{graphological identity}}. This emergent recognition practice creates a form of \textit{pseudo-anonymity}, where users remain unnamed, but not necessarily unidentifiable.
This tension is especially concerning given current advances in handwriting recognition and identification. As discussed in the previous section (see section~\ref{sssec:accessibility}), even limited handwriting samples can serve as identifiers. Existing research demonstrates that machine learning systems can already match handwriting styles with high accuracy~\cite{graves2009novel, javidi2020deep, alkawaz2020handwriting, rehman2019automatic}. While users may believe their interactions on GOS are safely anonymous, this false sense of security may expose them to the risk of de-anonymization. 

Such risks are not merely speculative. Handwriting is commonly used for identity verification in official documents, including government\allowbreak \ issued IDs. If linked to external datasets or used maliciously, handwriting data from GOS could enable re-identification of users who shared sensitive, personal, or politically risky content. This raises ethical concerns particularly for vulnerable populations, including whistle-blowers, activists, and others who rely on anonymity for safety and expression ~\cite{10.1145/3234942,10.1145/3491102.3517484}.
We therefore emphasize the need for future research on users’ mental models of anonymity in graphonymous spaces. Understanding how users perceive the risks of handwriting-based identification, and how these perceptions align or diverge from technical realities, is critical for designing safer, privacy-conscious platforms that support expressive freedom without compromising user security.

\subsubsection{Adaptive Strategies and Conversational Complexity}

Although human behavior and communication are influenced by social norms, people's actions are not strictly dictated by these norms. Instead, individuals orient themselves to these norms based on the specific situation, applying them flexibly and creatively ~\cite{egbert2012introduction}. This perspective is evident in our analysis of GI. Our findings reveal that in a visually complex GOS, participants adapt their communication strategies to align with the platform's affordances such as visual cues, color coding, canvas hopping and non-verbal elements (arrows and underlining specific parts of a text), akin to gaze in face-to-face communication. In addition, unlike conventional CMC, where messages are forced into a strict linear order leading to overlapping speech and miscommunication in multi-group settings ~\cite{herring1999interactional}, users on the GOS take advantage of the platform's affordances and canvas-based interface to separate simultaneous conversations. As a result, they are able to maintain coherence even in complex multi-user settings. Paralinguistic cues, such as personalized drawings resemble the use of non-verbal tools like facial emojis in text-centric CMC that convey emotional states and enrich linguistic context ~\cite{bai2019systematic, pfeifer2022all}.
However, unlike emojis, which have fixed meanings, the drawings in GOS do more than just express emotions. They also serve as identity markers and are highly personalized, working alongside distinctive handwriting styles within an anonymous setting. Furthermore, the ability to monitor real-time GI highlights the immediacy of interactions. 

At the same time, these adaptive practices also introduce a layer of complexity that shapes communication on the GOS. The platform's unstructured design presents notable challenges. The absence of threading tools leads to disorganized sequence organization, especially in overlapping inputs from multiple users. These fragmented conversations are often worsened by delayed responses or minimal engagement, which disrupt the durability of discussions. Canvas hopping illustrates users' dynamic adaptation to cope with visual clutter and facilitate smoother conversations. However, this visual complexity leads to increased cognitive load on GOS users to maintain clarity. While using a stylus enhances engagement, concerns arise regarding legibility and readability. This is because users have varying writing styles, and there is currently no assistive feature to ensure legible exchanges. Overall, these insights deepen our understanding of how users navigate the intricacies of communication on platforms like GOS through creative strategies that capitalize on the platform's visual and spatial features. They also present design challenges to fully support creative exchanges without restricting users' freedom of expression. 

\subsection{Limitations, Open Questions, and Future Work}

Like all empirical studies, our study has its limitations. First, our analysis was based on 600 plus canvas pages from GOS, 20 semi-structured interviews, and 70 minutes of live observational data. While these sources provide valuable insights into emerging practices on the GOS, they only offer a limited perspective on the potential interactions. \add{Our interview sample was predominantly young adult women (73\% females, mean age 21.5), which may limit generalizability to other demographic groups.}
Second, our study focused primarily on English language content. Future work could explore graphonymous interaction in multilingual and cross-cultural contexts to examine how users from different cultural and linguistic backgrounds draw on multi-modal affordances to construct and interpret implicit meanings. Third, we did not directly compare GOS to mainstream social media platforms such as Facebook or Instagram, particularly in terms of user experience and coordination practices. Comparative studies could reveal how socio-technical dynamics differ across platforms, especially between linear, profile-based systems and freeform, canvas-based ones. Lastly, while we included interview questions about trust dynamics (section~\ref{interviewsonGOS}), participants provided limited elaboration on this dimension, preventing systematic thematic analysis. 
This study also brings to light several open research questions. \textbf{\textit{From an accessibility standpoint}}:  
\textit{How can accessibility systems interpret and support spatial, non-linear, and stylistic elements of visual interaction in graphonymous environments?}\textbf{\textit{ From an interactional perspective}}:  
\textit{How can future systems support conversational clarity, turn-taking, and mutual awareness in visually rich, non-threaded spaces without constraining user creativity?}\textbf{\textit{ From a security perspective:}}  
\textit{How do users understand the risks of de-anonymization through handwriting in anonymous platforms, and how can such risks be mitigated without compromising expressive freedom?}~\textbf{\textit{ From the standpoint of trust:}}  
\textit{\add{How do users develop interpersonal trust across multiple sessions? and Does graphological recognition create stronger trust bonds than text-based anonymity?}}.

Additionally, the issue of moderation and governance in GOS-like platforms remains an open challenge. In shared, anonymous visual spaces, users can overwrite or delete each other's contributions with little trace or accountability. We observed cases where conversations were disrupted by unsolicited erasure, \textbf{\textit{raising questions about authorship, ownership, and resilience in collaborative environments}}: \textit{What mechanisms can preserve the integrity of conversations or artworks in GOS, particularly in the face of intentional deletion or interference? }
\textit{How can moderation systems be designed for co-created, short-lived, and visually intertwined content without relying on identity markers or centralized control?}

In light of these questions, we propose several directions for future design. Optimizing GOS's interactional affordances could support smoother participation and reduce friction. For instance, automatic canvas resizing could reduce clutter, while visual labels or conversation tags could help users navigate parallel threads. Handwriting correction or stylisation tools could also improve legibility for diverse contributors. Moderation tools such as non-invasive visual history, soft locks on active contributions, or undo options might also support respectful co-creation, while preserving GOS's open and anonymous ethos. These limitations and questions collectively demonstrate that GOS and GI represent a promising new space for studying and designing expressive, anonymous, and collaborative interactions. However, they also raise important ethical, technical, and interactional challenges for future HCI research. 

\section{Conclusion}

Online social networks have historically relied on text-based communication, with substantial research devoted to understanding interaction, creativity, and engagement in these environments. Yet alternative modalities such as digital handwriting and drawing have received far less attention. This study introduces \textit{Graphonymous Interaction} (GI), a novel mode of anonymous online communication based on digital handwriting and drawing, situated within collaborative environments we term \textit{Graphonymous Online Space} (GOS).

Our findings reveal that GOS fosters creativity, emotional expression, and social connection through its uniquely visual and personalized forms of interaction. Users employ handwriting and drawing not only for communication but also to express their identity, convey affect, and build interpersonal relationships that go beyond the limitations of text. Unlike traditional messaging systems, GOS enables fluid, asynchronous interaction with fewer overlaps and conversational breakdowns, all thanks to users’ real-time visual awareness of each other’s contributions.

Beyond emotional and relational expression, GOS also supports collaborative creativity and intellectual engagement. Activities such as drawing games, shared canvases, and co-authored artworks serve as sites for playful problem-solving and collective meaning-making. These insights suggest that graphonymous online spaces have significant potential to re-imagine social media. They offer expressive, egalitarian, and less identity- and metric-driven environments. By identifying and characterizing GI, and analyzing the practices that emerge in GOS, this research contributes to a broader understanding of anonymous, creative, and visual forms of social interaction. Furthermore, it opens new directions for designing expressive, accessible, and secure online platforms.

\begin{acks}
    During the preparation of this manuscript, we employed generative AI tools, including Paperpal and ChatGPT-5, strictly as aids to enhance grammar, spelling, and overall readability. The core ideas, data analyses, and content are solely the authors' original work.
\end{acks}

\bibliographystyle{ACM-Reference-Format}
\bibliography{GOS}

\end{document}